\documentclass[reprint,superscriptaddress, aps, prb,%rsi,
amsmath,amssymb,longbibliography]{revtex4-2}
\usepackage[T1]{fontenc}
\usepackage{graphicx}
\usepackage{dcolumn}
\usepackage{bm}
\usepackage{color}
\usepackage{textgreek}
\usepackage{braket, comment}
\usepackage{siunitx}

\begin{document}
\title{Microwave-free imaging magnetometry with nitrogen-vacancy centers in nanodiamonds at near-zero field}

\author{Saravanan Sengottuvel} \thanks{These authors contributed equally to this work}
\affiliation{Jagiellonian University, Doctoral School of Exact and Natural Sciences, {\L}ojasiewicza 11, 30-348 Krak\'{o}w, Poland}\email{saravanan.sengottuvel@doctoral.uj.edu.pl}
\affiliation{%
Institute of Physics, Jagiellonian University, {\L}ojasiewicza 11, 30-348 Krak\'{o}w, Poland}%

\author{Omkar Dhungel} \thanks{These authors contributed equally to this work}
\affiliation{Helmholtz-Institut Mainz, 55099 Mainz, Germany}

\affiliation{Johannes Gutenberg-Universit{\"a}t Mainz, 55128 Mainz, Germany}
\author{Mariusz Mrózek }
\affiliation{%
Institute of Physics, Jagiellonian University, {\L}ojasiewicza 11, 30-348 Krak\'{o}w, Poland}%

\author{Arne Wickenbrock}
\affiliation{Helmholtz-Institut Mainz, 55099 Mainz, Germany}
\affiliation{Johannes Gutenberg-Universit{\"a}t Mainz, 55128 Mainz, Germany}
\affiliation{GSI Helmholtzzentrum f{\"u}r Schwerionenforschung GmbH, 64291 Darmstadt, Germany}

\author{Dmitry Budker}
\affiliation{Helmholtz-Institut Mainz, 55099 Mainz, Germany}
\affiliation{Johannes Gutenberg-Universit{\"a}t Mainz, 55128 Mainz, Germany}

\affiliation{GSI Helmholtzzentrum f{\"u}r Schwerionenforschung GmbH, 64291 Darmstadt, Germany}

\affiliation{Department of Physics, University of California, Berkeley, California 94720-7300, USA}

\author{Wojciech Gawlik }
\affiliation{%
Institute of Physics, Jagiellonian University, {\L}ojasiewicza 11, 30-348 Krak\'{o}w, Poland}%

\author{Adam M. Wojciechowski}\email{a.wojciechowski@uj.edu.pl}
\affiliation{%
Institute of Physics, Jagiellonian University, {\L}ojasiewicza 11, 30-348 Krak\'{o}w, Poland}%
\date{\today}

%########### ABSTRACT #################
\begin{abstract} 
Magnetometry using Nitrogen-Vacancy (NV) color centers in diamond predominantly relies on microwave spectroscopy. However, microwaves may hinder certain studies involving biological systems or thin conductive samples. This work demonstrates a wide-field, microwave-free imaging magnetometer utilizing NV centers in nanodiamonds by exploiting the cross-relaxation feature near zero magnetic fields under ambient conditions without applying microwaves. For this purpose, we measure the center shift, contrast, and linewidth of the zero-field cross-relaxation in 140 nm nanodiamonds drop-cast on a current-carrying conductive pattern while scanning a background magnetic field, achieving a sensitivity of 4.5\,\si{\micro\tesla/\sqrt{\hertz}}. Our work allows for applying the NV zero-field feature in nanodiamonds for magnetic field sensing in the zero and low-field regimes and highlights the potential for microwave-free all-optical wide-field magnetometry based on nanodiamonds.
\end{abstract}
\maketitle 

%########### INTRODUCTION #################
\section{Introduction}
The application of negatively charged nitrogen-vacancy (NV$^{-}$) color centers in diamond \cite{gali2019ab} has become the focal point for sensing physical parameters such as temperature, pressure, magnetic and electric fields at the nanoscale, even under challenging environmental conditions \cite{fu2020sensitive, wang2015high, ho2021recent, kuwahata2020magnetometer, dolde2011electric}. NV centers have demonstrated remarkable sensitivities in recent years, reaching 0.5\,\si{\pico\tesla/\sqrt{\hertz}} for ensembles at room temperature \cite{barry2023sensitive}. The use of NV centers for magnetometry allows the exploration of magnetic fields generated by various samples, including biological systems, magnetic materials, and current-carrying wires \cite{barry2016optical, LenzPRA2021, nowodzinski2015nitrogen, levine2019principles}. NV magnetometry leverages the magnetically sensitive microwave transitions via the Optically Detected Magnetic Resonance (ODMR), which relies on detecting the spin-dependent fluorescence while applying microwaves to the optically pumped NV centers. 

Many applications of magnetic measurement using NV centers are based on microwave-induced ODMR. However, these techniques can be challenging when studying systems where external magnetic fields and high-power microwaves could interfere with the sample, where it is difficult to implement such control, or where microwaves are invasive, such as in the case of high-transition-temperature (T$_{c}$) superconductors, samples for zero- to ultra-low-field nuclear magnetic resonance (ZULF NMR) measurements, thin conductive materials, and biological samples \cite{BLANCHARD2021106886, simpson2016magneto}. 

To address these problems, microwave-free protocols have been developed. They exploited energy-level crossings of NV centers at different non-zero magnetic fields, notably the level anti-crossing within the triplet ground state at 102.4\,\si{\milli\tesla}, as well as the NV-P1 crossing at 51.2\,\si{\milli\tesla} \cite{doi:10.1063/1.4960171, https://doi.org/10.1002/qute.202000037, 10.1063/5.0059330, PhysRevApplied.13.044023, rebeirro2024microwave}. However, in addition to the nonzero-field crossings, the electronic sublevels of the NV center cross also at a zero \textit{B}-field \cite{Rondin2014, mrozek2015EPJ, blanchard2007zero}. Each level crossing reflects the related degeneracy in the system that makes it very prone to perturbations that may mix the quantum states and alter their populations, which is known as cross-relaxation \cite{Oort1989PRB}. 

Alternatively, various zero-field magnetic measurement strategies using NV centers have been proposed \cite{kong2018nanoscale, wang2022zero, PhysRevApplied.11.064068, lenz2021}, which involve microwaves. However, these techniques can be challenging for studying systems where high-power microwaves and a high external magnetic field might interfere with the sample of interest. To address this, an NV magnetometry scheme at near zero fields without microwave radiation has been proposed \cite{PhysRevA.94.021401, Anishchik2015, AkhmedzhanovPRA2017}. This technique involves measuring the magnetic field by observing cross-relaxation features in the fluorescence spectrum of NV centers at near zero \textit{B}-fields. Let us assume that the measured magnetic field is the sum of an unknown constant field and a scanned field. In this scenario, the zero-field cross-relaxation feature position depends on the constant field and can be used to infer the unknown magnetic field.

This work focuses on the implementation of a practical NV magnetometer utilizing the zero-field cross-relaxation feature. Specifically, we present a microwave-free magnetometer that leverages the cross-relaxation feature at zero-field to map magnetic fields above a current-carrying cross-pattern coated with NV nanodiamonds. The use of nanodiamonds facilitates imaging on arbitrarily shaped surfaces. Our findings highlight the potential of zero-field magnetometry for real-world applications.

%########### NEAR ZERO_FIELD MICROWAVE MAGNETOMETRY #################

\section{Zero-field microwave-free NV magnetometry}
Numerous studies have been carried out recently to understand NV cross-relaxation characteristics near zero-field. Although not all aspects of longitudinal relaxation ($T_1$) in dense ensembles are fully understood \cite{cambria2023temperature}, one hypothesis involves ``fluctuator'' defects, which depolarize nearby NVs, leading to depolarization of the entire sample via dipolar interactions \cite{PhysRevLett.118.093601}. In addition, there is also a relaxation mechanism associated with local electric fields and the interaction between NV centers oriented similarly and differently, which is particularly important in high-density samples \cite{pellet2023relaxation}. The zero field cross-relaxation feature and additional features observed in the presence of a bias field have been systematically examined as a function of NV density, magnetic field orientation, and temperature \cite{dhungel2023zero}. Similar characteristics have also been identified in nanodiamonds, and their behavior has been investigated in terms of nanodiamond size and light intensity. Specifically, the authors of Ref. \cite{Dhungel:24} investigated how variations in NV density, orientation, nanodiamond size, and light intensity influence observed characteristics, providing a significant understanding of the interplay of these factors in the studied phenomena.

Magnetometry with nanodiamonds offers advantages over other NV-diamond sensors. Conventional NV magnetometry often involves the use of bulk diamond or thin NV layers or employs scanning diamond tips to reduce the standoff distance to the sample, typically down to several nanometers for AFM-type sensors, to achieve high sensitivity and spatial resolution when mapping magnetic fields \cite{huxter2022scanning}. However, these methods have limitations as they require a smooth sample surface and proximity of the sample to the measurement diamond. In contrast, the small size of nanodiamonds allows them to be bonded to irregular material surfaces, including fiber tips and living cells \cite{filipkowski2022volumetric, mi2021quantifying}.

In this study, we used the "salt and pepper" approach \cite{Dhungel:24} using randomly oriented nanodiamonds containing NV centers to create a cost-effective and straightforward wide-field magnetometer without microwaves for near-zero-field measurements. The sensitivity of these NV-ND magnetometers can reach below the \si{\nano\tesla}/$\sqrt{\textrm{Hz}}$ level \cite{chen2022nanodiamond}, with typical per-pixel sensitivities in the case of imaging devices of a few \si{\micro\tesla}/\si{\sqrt{\hertz}} \cite{sengottuvel2022wide}. The biocompatibility of NDs allows for their use in biomedical applications, such as magnetic resonance imaging (MRI) of the brain and other organs \cite{devience2015nanoscale}. Furthermore, the small size of NDs enables nanoscale spatial resolution, making them ideal for imaging the magnetic fields of biological structures at the cellular and subcellular levels \cite{Barbiero20}. The utilization of nanodiamonds presents a balanced compromise between sensitivity, resolution, and manufacturing costs. Another significant advantage of our method is the implementation of wide-field magnetic imaging without microwaves, eliminating all the effects of microwaves, enabling real-time parallel mapping of the magnetic field across a large field of view, and reducing measurement time.
\onecolumngrid

\begin{figure*}[t]
    \includegraphics[width=0.8\textwidth]{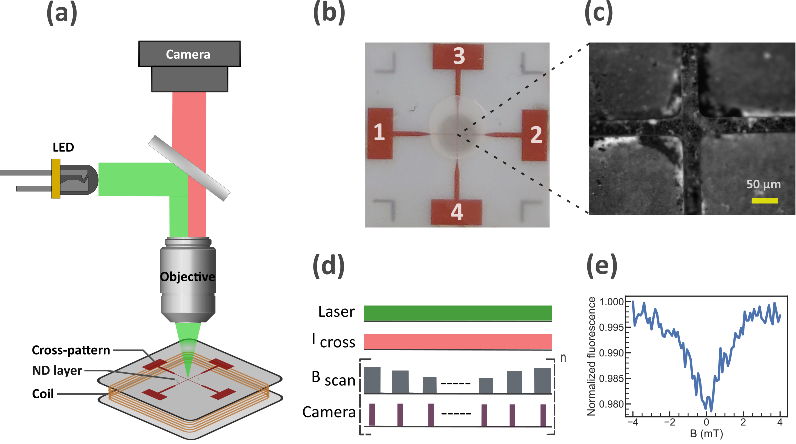}
    \caption{(a) schematic of the wide-field magnetic imaging setup, (b) the conductive cross pattern embedded in a transparent substrate attached to a PCB with NDs deposited at the interjunction, (c) a single-shot microscopic fluorescence image of NV centers with the cross pattern in the background, (d) timing of the measurement and imaging protocol, (e) near-zero-field fluorescence spectra of the nanodiamonds.}
    \label{fig:setup}
\end{figure*}
\vspace*{\columnsep}
\twocolumngrid
%########### EXPERIMENTAL SETUP #################
\section{Experimental setup}
The diamonds used in this work are commercially available 140-\,\si{\nano\meter} carboxylated fluorescent nanodiamonds by Adamas Nanotechnologies. A suspension of these diamonds in deionized (DI) water is deposited onto a transparent polyethylene terephthalate (PET) substrate, which has a thickness of 0.11 mm, using the drop-casting technique. The sample is then allowed to air-dry for 15 to 30 minutes. On the reverse side of the PET substrate, a copper cross-pattern, with wires 65\,\si{\micro\meter} wide, is printed and subsequently fixed onto a printed circuit board (PCB). This cross-pattern comprises four individual arms (labeled 1 to 4 in Fig. Fig. \ref{fig:setup}b)) that intersect at the center. The pattern is connected to a power source, allowing the current to be driven through any specific path (e.g., 1-2, 1-4) across the pattern.

For NV magnetometry described here, a home-built wide-field fluorescence microscope was used, as shown in  (Fig.\,\ref{fig:setup}a). The nanodiamond sample was illuminated with 60-70\,\si{\milli\watt} of 532 \si{\nano\meter} green light from an LED, Thorlabs M530L4). The light beam was collimated with an aspheric condenser lens (focal length 50\,\si{\milli\meter}) and guided with a dichroic mirror to the back focal plane of a 40x Olympus microscope objective with a numerical aperture of 0.65. The pump beam focused with the objective illuminates the nanodiamonds. The red fluorescence emitted from the NV centers is collected using the same microscope objective and imaged with a Basler camera equipped with a 12-bit Sony CMOS sensor. In our imaging system,  the camera field of view (FOV) is 2448 $\times$ 2048 pixels which corresponds to an area of 384\,\si{\micro\meter}$\times$\,321\,\si{\micro\meter} on the nanodiamond layer with 0.15\,\si{\micro\meter}$\times$ 0.15\,\si{\micro\meter} per pixel. A square DC current-carrying coil (with N = 20 turns and length $R_{coil}$ =  50\,\si{\milli\meter}) creates a bias field perpendicular to the NV imaging plane.

The NV magnetometer was calibrated using a test magnetic field. The magnetic field of the current-carrying coil was cross-verified by measuring the Zeeman splitting along the quantization axis (\{111\} crystallographic directions) of one of the NV orientations in a 100-cut thin monocrystal diamond sample. The estimated magnetic field value from NV ODMR splitting is consistent with the applied test magnetic field calculated based on the coil current $I$ and its geometry.
 
%########### RESULTS AND DISCUSSIONS #################
\section{Results and discussions}
\subsection{Imaging of static magnetic fields using nanodiamonds}\label{characterization}

In this section, we describe the methodology to image and map the magnetic field generated with a cross pattern for various current intensities and paths. As illustrated in Fig.\,\ref{fig:setup}(d), both the laser and the DC remain constant throughout the measurement. The background magnetic field applied perpendicular to the plane of the ND layer is systematically scanned in steps from -4.0\,\si{\milli\tesla} to +4.0\,\si{\milli\tesla} with the corresponding images captured sequentially. 

Before the current measurements, we experimentally verified the presence of the zero-field feature in the NDs in the absence of current in the cross. Next, the non-zero current in the cross-pattern was applied in two specific directions: i) path 3 to 4 and ii) path 1 to 4. We took the measurement several times and averaged the resulting images to improve the signal-to-noise (SNR) ratio. For a given current path in the cross, we scan the magnetic field and record the fluorescence image of the ND layer. From the acquired image data, we then extract the zero-field cross-relaxation spectrum for each pixel of the camera. To reduce the relative noise of each pixel and the amount of data to be processed, we binned 16$\times$16 pixels, thus reducing the total pixel number by a factor of 256. The extracted spectrum is then fitted with a Gaussian function (\ref{eq:gaussian}) to estimate the shift of the center field ($\Delta B$), full-width at half maximum (\textit{w}), and the contrast (C) of the zero-field cross-relaxation feature.
\begin{equation}
    f(x; \Delta B, w) = y_{0} + C \cdot \frac{1}{w\sqrt{2\pi}} e^{-\frac{(x - \Delta B)^2}{2w^2}}\,.
\label{eq:gaussian}
\end{equation}
Here, the parameters $y_{0}$, $C$, $\Delta B$, and \textit{w} are, respectively, the offset, contrast, center, and width of the distribution. An additional magnetic field generated by the cross-pattern in both parallel and perpendicular directions to the scanned field affects the parameters $\Delta B$, $C$, and\textit{w} of the zero-field feature. The shift $\Delta B$ is equal to the strength of the \textit{B}-field generated above the cross-pattern. Quantifying these parameters spatially provides information about the field generated by the cross-pattern.

\begin{figure*}[t!]
\centering
    \includegraphics[width=0.9\textwidth]{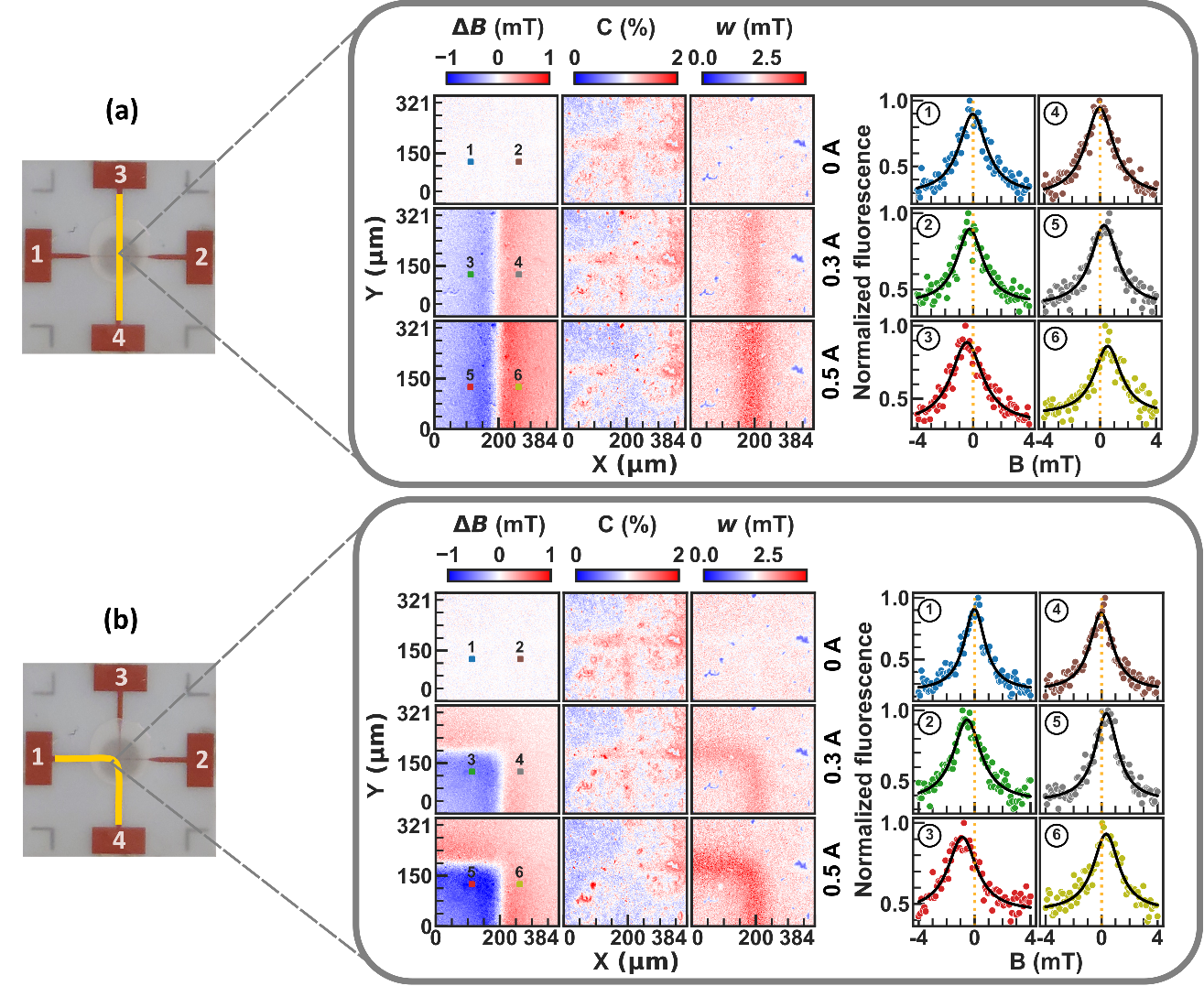}
    \caption{Maps of the shift $\Delta B$, contrast C, and width \textit{w} of zero-field spectra for two current paths in the cross pattern with varying current intensities of 0\,A, 0.3\,A, and 0.5\,A. (a)The current path is from point 3 to 4; (b) The current path is from point 1 to 4.The plots to the right of the maps are the single-pixel zero-field spectra extracted from the binned image data. The pixel locations are colored in the width maps.} 
\label{fig:maps}
\end{figure*}

The Gaussian function is then fitted to the extracted spectrum to obtain the $\Delta B$, $C$, and \textit{w} values for each pixel. Using these estimated values, we create $\Delta B$, $C$, and \textit{w} maps for the entire field of view (FOV) of the camera under two different current paths at 0\,A, 0.3\,A, and 0.5\,A, as shown in Fig-\ref{fig:maps}.

As expected, when there is no current (0A) flowing through the conductive cross-pattern, we see no visible trace of the pattern in the $\Delta B$ map, which is expected. However, when the current is non-zero (0.3\,A and 0.5\,A), the arms of the cross pattern (the central white region), through which the current flows, become visible. Notably, shift $\Delta B$ above the cross-pattern is zero, regardless of the current path, from 3 to 4 or 1 to 4. This is because the magnetic field generated by the conductive cross-pattern is perpendicular to the scanned field and does not affect it. 
\onecolumngrid

\vspace{\columnsep}
\twocolumngrid
The strength of the magnetic field decreases with distance from the center of the wire, which is observable in the shift maps. The color scheme on the center field shift map indicates that the magnetic field generated by the wire is either parallel (red) or anti-parallel (blue) to the scanned field. The thickness of the wire estimated from the map agrees with the actual dimensions. The colored squares on both the left and right sides of the pattern indicate pixel locations, and the corresponding spectra are shown in Figure 2 to the right of the maps. The spectra extracted from the binned image data visualize the center shift when a current is applied; the peak shift increases with a higher current and reverses when the current is reversed.

The contrast and width maps provide complementary information to the shift maps. At 0\,A, with no current flowing through the conductive cross-pattern, we observe a maximum contrast of the zero-field feature of around 1\% and a minimum width of 2 mT for nanodiamonds throughout the FOV. As the current increases, the overall contrast of the zero-field feature reduces, and its width increases, as evident from the colormaps at 0.3\,A and 0.5\,A. This occurs because the magnetic field from the cross-pattern tunes the electron spin splitting of NV centers into resonance with many surrounding NV centers that are randomly oriented. This results in less-enhanced cross-relaxation through magnetic dipole interactions with neighboring NV centers, leading to higher emissions. At 0\,A, unlike in the $\Delta B$ maps, we observe a trace of the conductive cross-pattern in the contrast and width maps, though it is less pronounced in the latter. This trace is due to the reflective nature of the copper structure.
Additionally, Fig-\ref{fig:maps} shows a non-uniformity of contrast and width around the cross-pattern, where the current is maximum at the center and decreases toward the edges of the maps. This suggests the presence of a non-uniform NV layer and nanodiamond clusters, possibly resulting from the aggregation of nanodiamonds during deposition on the PET surface. Also, "oasis-like" structures are seen in the width map, which may be due to the failure of the fitting function to model the data accurately.

The center field shift ($\Delta B$) above the cross pattern is zero when the direction of the current is from 3 to 4 and from 1 to 4. 
\begin{figure}[h!]
     \centering  
     \includegraphics[width=\columnwidth]{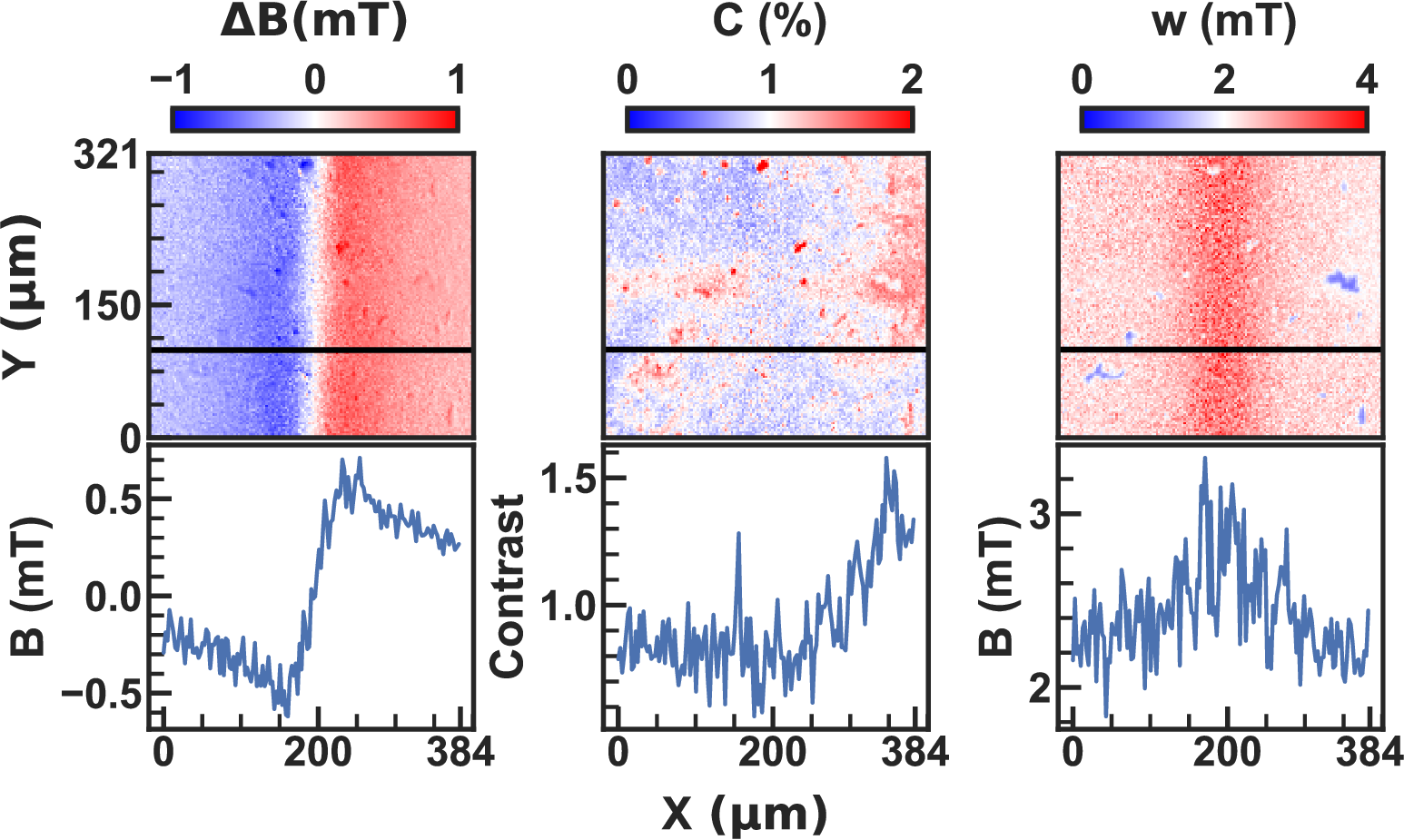} 
     \caption{The first row shows maps of the center shift $\Delta B$, contrast C, and width \textit{w} of the zero-field feature for I = 0.5\, A. The second row shows the $z$ component of the magnetic field, contrast, and width as a function of distance from the cross for a specific row of pixels marked by a black horizontal line in the upper row images.} %\DB{No, this is NOT what these plots show! Maybe the first column can be related to the Z component of the field...} \DB{Need to explain the horizontal black line on the upper plots.}}
     \label{fig:singlepixel}
 \end{figure} This is because the magnetic field generated by the current-carrying pattern is perpendicular to the scanned field, broadening the zero-field feature and decreasing the observed contrast. Additionally, the strength of the field decreases with the distance from the center of the wire, which can be observed in center shift maps. The color code (blue to red) on the center field shift map indicates that the magnetic field generated by the wire is either parallel or anti-parallel to the scanned field. Contrast and width maps offer complementary information about the broadening when the magnetic field is perpendicular to the scanning background field.
\begin{figure}[h!]
     \centering
     \includegraphics[width=0.65\columnwidth]{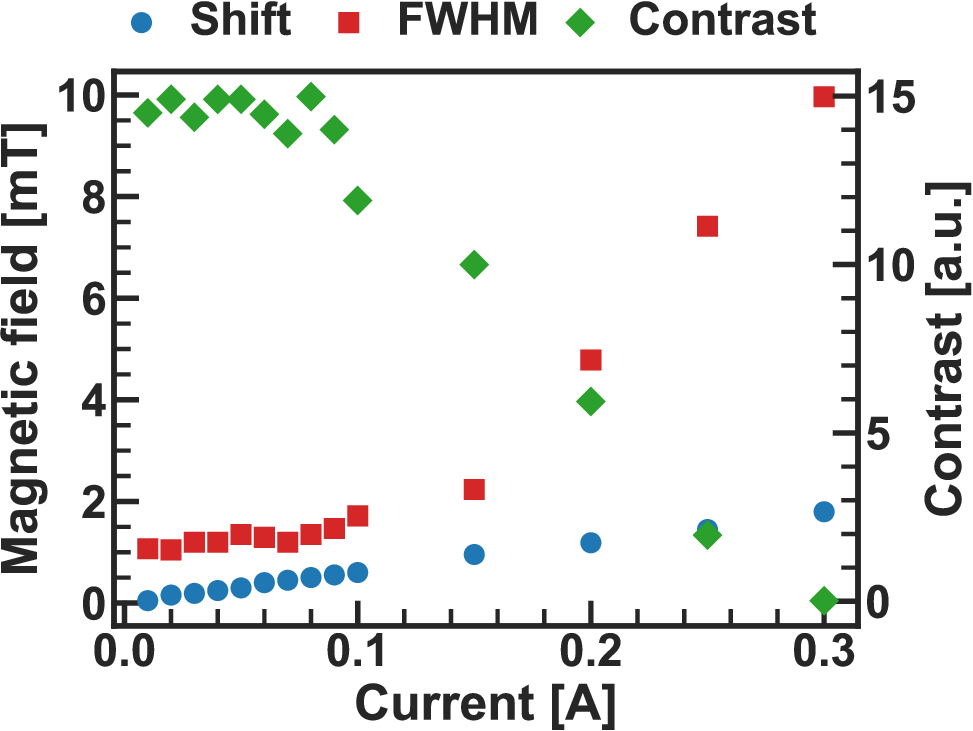}     
     \caption{Change in the Gaussian fit parameters as a function of applied current (shift and width correspond to the left scale, Contrast to the right scale).}
     \label{fig:paramVScurr}
 \end{figure}

Figure \ref{fig:singlepixel} shows the map of the zero field parameters and cross-sectional plots of the parameters for an applied current of 0.5\,A. The plot observed for $\Delta B$, obtained by selecting one row of the image data perpendicular to the current carrying wire, shows that the maximum (minimum) shift occurs when the applied test field is parallel (anti-parallel) to the scanning background field and decreases with distance from the center. The shift is zero in the center of the wire, while the contrast\,($C$) is minimum and width (\textit{w}) is maximum there.

Our measurements show that the changes in $\Delta B$, $C$, and \textit{w} are proportional to the current applied to the wire in the relevant distance ranges. Figure\,\ref{fig:paramVScurr} shows that the fit parameters depend linearly on the current flowing in the cross pattern for the center shift, the width, and the contrast above 0.1\,A.

\subsection{Numerical simulations}
\begin{figure}[h]
     \centering
     \includegraphics[width=0.9\columnwidth]{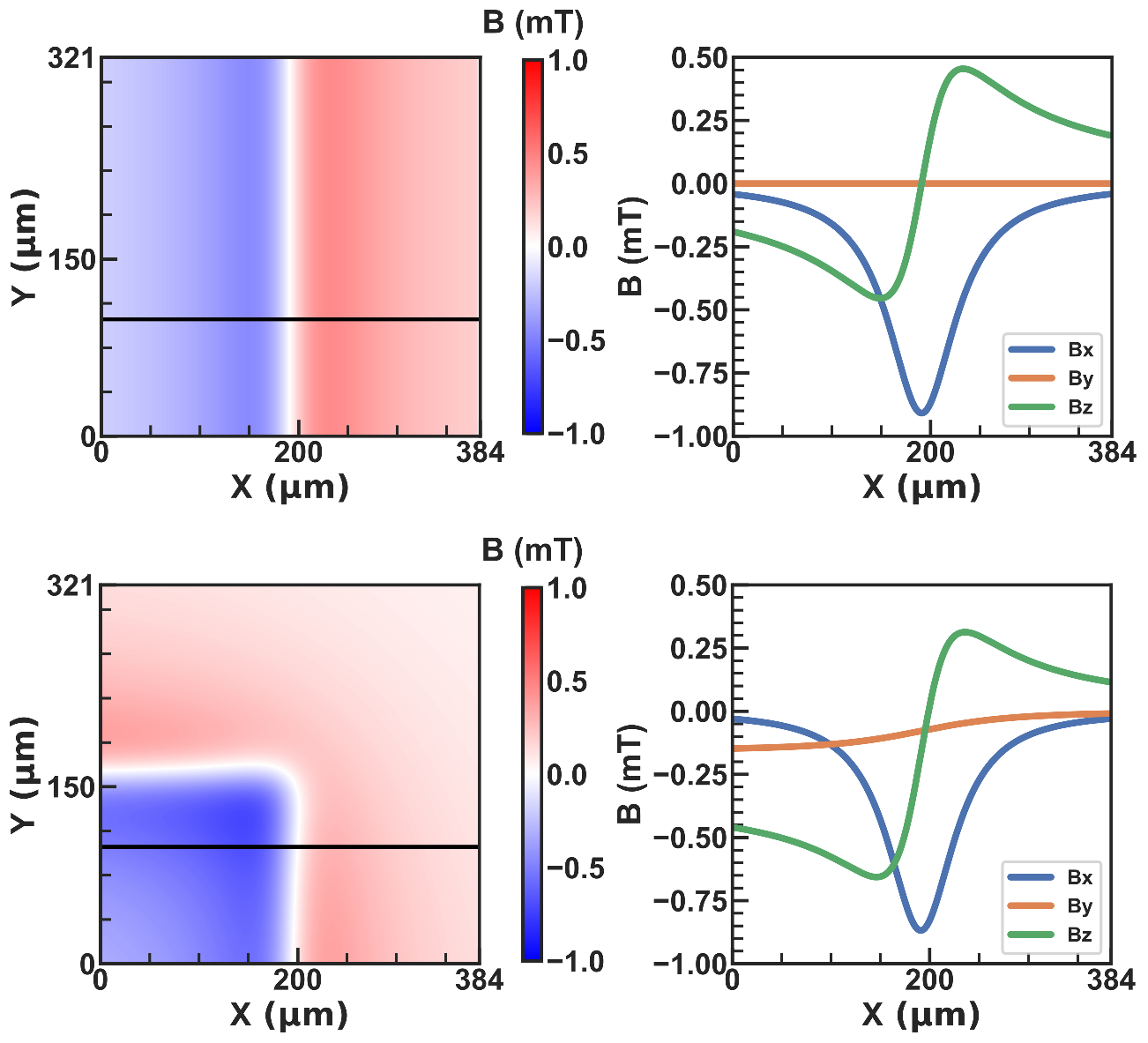}
     \caption{Numerical simulation of the \textit{B}-field distribution at a distance of 0.11 mm from the surface of the cross pattern with an applied current of 0.5\,A in two specific directions (a) path 3 to 4 and (b) path 1 to 4. The plots adjacent to the maps show the components of the \textit{B}-field along the black line.}
    \label{fig:simulation}
 \end{figure} 
To cross-validate our microwave-free magnetometry scheme, we performed a numerical simulation. Numerical integration was performed using the open-source Python package Magpylib \cite{Ortner2020}. The simulated \textit{B}-field distributions for two different current paths are shown in Fig.\,\ref{fig:simulation}. The current-induced fields of currents are directly derived from the Biot-Savart law. The magnetic field distribution map in the $xy$ plane was calculated at 0.11\,\si{\milli\meter} along the Z-axis from the cross-structure to account for the thickness of the substrate. As can be seen, there is a good agreement between the simulation and our measurements.
 
\subsection{Single-pixel magnetic field sensitivity}
The photon shot-noise limited magnetic field sensitivity of an NV ensemble magnetometer (similarly to that of a continuous-wave ODMR imaging protocol) is given by \cite{Barry2020aps, Dreau2011PhysRev}:
\begin{equation}
    \delta{B} = P_{F}\frac{h}{g\mu_{B}}\frac{\Gamma}{C\cdot \sqrt{I_{0}}}
    \label{eqn:Bsens}
\end{equation}

Here $\Gamma$ is the FWHM resonance linewidth, $C$ is the contrast, $P_{F}\,\approx$\,0.70 for a Gaussian-shaped resonance peak, $I_{0}$ is the photon detection rate. Here, the detected photon rate is defined in terms of the number of photon counts per pixel per second, taking into account the sensor quantum efficiency. Equation\,\ref{eqn:Bsens} can be applied similarly to estimate the NV magnetic field sensitivity using the zero-field feature since the feature width and the contrast primarily depend on the number of NV centers in the nanodiamonds. Due to the random orientation of NV centers in NDs and the variation in the number of nanodiamonds per pixel, we estimate that the mean per-pixel sensitivity of our ND magnetometer is approximately 4.5\,\si{\micro\tesla}/\si{\sqrt{\hertz}} per \si{\micro\meter^{2}} ($\approx$ $9 \times 10^6$ counts per second for an effective sample area of 0.15\,\si{\micro\meter^{2}}). %Moreover, for comparing with the sensitivity of other NV magnetometers, we estimated the area-normalized sensitivity and the value is 4\,\si{\micro\tesla}\si{\micro\meter}/\si{\sqrt{\hertz}}. \SA{I will cross-check these values}%\OM{I would mention values of all parameters so that people can get some feeling.}
The estimated sensitivity of our current imaging instrument corresponds to the field sensed by a 140\,\si{\nano\meter} nanodiamond sitting at a distance of 110um above the cross-pattern. In our experiment, this sensitivity was sufficient to detect \textit{B}-fields ($<$ 1\,\si{\milli\tesla}) when applying a few mA of current. The sensitivity can be further improved by reducing the stand-off distance between the NV layer and the current source, which would allow the detection of smaller magnetic fields and increase the precision of source localization.

%########### CONCLUSIONS #################
\section{Conclusions}
In this study, we have shown that the zero-field cross-relaxation feature can be effectively used for imaging and mapping the magnetic field distribution. As a proof-of-concept potential application, we successfully visualized the magnetic field produced by a current-carrying wire pattern using NV ensembles in nanodiamonds. 

In this work, we characterized three key parameters of the zero-field feature: center, width, and contrast for different current intensities. Additionally, we validated our experimentally observed magnetic field distribution of the cross-pattern through numerical simulations. In particular, the feature exhibits a broader linewidth ($\approx$ 2.0\,\si{\milli\tesla}) and a lower contrast of approximately 1--2\% in NDs compared to some single crystal diamond samples with high NV concentrations ($\approx$ 0.2\,\si{\milli\tesla}) linewidth and $\approx$ 4\% contrast \cite{dhungel2023zero}.

The fact that the zero-field technique does not require a microwave or precise orientation of a known magnetic field makes it unique among other NV magnetometers. The zero-field technique may find applications in biosensing since the lack of microwave field is a significant advantage when studying cells, and nanodiamonds are already readily used as trackers for fluorescence microscopy. It could also be used for real-time magnetometry in living cells with diamonds without having to constantly calibrate the diamond axes using ODMR and other microwave-based techniques. Thanks to the application of nanodiamond coatings, the method can be very useful for imaging of arbitrary-shaped surfaces.

 The measured sensitivity with the zero-field technique is comparable to that obtained with the continuous-wave ODMR (cw-ODMR) and GSLAC techniques. We stress that the zero-field microwave-free technique is not optimal since the technique used here is analogous to cw-ODMR, which is not the most sensitive DC magnetometer. Further, we aim to improve the sensitivity of our magnetometer by optimizing the experimental setup. However, how to improve its sensitivity with the material engineering of nanodiamonds remains unclear. 

\section*{Acknowledgements} 
 This work was supported by the European Commission’s Horizon Europe Framework Program under the Research and Innovation Action MUQUABIS GA no. 101070546, by the German Federal Ministry of Education and Research (BMBF) within the Quantumtechnologien program (DIAQNOS,
project no. 13N16455) and by the German DFG, Project SFB 1552 ``Defekte und Defektkontrolle in weicher Materie”. The study was carried out using research infrastructure purchased with the funds of the European Union in the framework of the Smart Growth Operational Programme, Measure 4.2; Grant No. POIR.04.02.00-00-D001/20, “ATOMIN 2.0 - ATOMic scale science for the INnovative economy and was also funded by the National Science Centre, Poland grant number 2020/39/I/ST3/02322.

\bibliography{main.bib}

%apsrev4-2.bst 2019-01-14 (MD) hand-edited version of apsrev4-1.bst
%Control: key (0)
%Control: author (8) initials jnrlst
%Control: editor formatted (1) identically to author
%Control: production of article title (0) allowed
%Control: page (0) single
%Control: year (1) truncated
%Control: production of eprint (0) enabled
\begin{thebibliography}{44}%
\makeatletter
\providecommand \@ifxundefined [1]{%
 \@ifx{#1\undefined}
}%
\providecommand \@ifnum [1]{%
 \ifnum #1\expandafter \@firstoftwo
 \else \expandafter \@secondoftwo
 \fi
}%
\providecommand \@ifx [1]{%
 \ifx #1\expandafter \@firstoftwo
 \else \expandafter \@secondoftwo
 \fi
}%
\providecommand \natexlab [1]{#1}%
\providecommand \enquote  [1]{``#1''}%
\providecommand \bibnamefont  [1]{#1}%
\providecommand \bibfnamefont [1]{#1}%
\providecommand \citenamefont [1]{#1}%
\providecommand \href@noop [0]{\@secondoftwo}%
\providecommand \href [0]{\begingroup \@sanitize@url \@href}%
\providecommand \@href[1]{\@@startlink{#1}\@@href}%
\providecommand \@@href[1]{\endgroup#1\@@endlink}%
\providecommand \@sanitize@url [0]{\catcode `\\12\catcode `\$12\catcode `\&12\catcode `\#12\catcode `\^12\catcode `\_12\catcode `\%12\relax}%
\providecommand \@@startlink[1]{}%
\providecommand \@@endlink[0]{}%
\providecommand \url  [0]{\begingroup\@sanitize@url \@url }%
\providecommand \@url [1]{\endgroup\@href {#1}{\urlprefix }}%
\providecommand \urlprefix  [0]{URL }%
\providecommand \Eprint [0]{\href }%
\providecommand \doibase [0]{https://doi.org/}%
\providecommand \selectlanguage [0]{\@gobble}%
\providecommand \bibinfo  [0]{\@secondoftwo}%
\providecommand \bibfield  [0]{\@secondoftwo}%
\providecommand \translation [1]{[#1]}%
\providecommand \BibitemOpen [0]{}%
\providecommand \bibitemStop [0]{}%
\providecommand \bibitemNoStop [0]{.\EOS\space}%
\providecommand \EOS [0]{\spacefactor3000\relax}%
\providecommand \BibitemShut  [1]{\csname bibitem#1\endcsname}%
\let\auto@bib@innerbib\@empty
%</preamble>
\bibitem [{\citenamefont {Gali}(2019)}]{gali2019ab}%
  \BibitemOpen
  \bibfield  {author} {\bibinfo {author} {\bibfnamefont {{\'A}.}~\bibnamefont {Gali}},\ }\bibfield  {title} {\bibinfo {title} {Ab initio theory of the nitrogen-vacancy center in diamond},\ }\href@noop {} {\bibfield  {journal} {\bibinfo  {journal} {Nanophotonics}\ }\textbf {\bibinfo {volume} {8}},\ \bibinfo {pages} {1907} (\bibinfo {year} {2019})}\BibitemShut {NoStop}%
\bibitem [{\citenamefont {Fu}\ \emph {et~al.}(2020)\citenamefont {Fu}, \citenamefont {Iwata}, \citenamefont {Wickenbrock},\ and\ \citenamefont {Budker}}]{fu2020sensitive}%
  \BibitemOpen
  \bibfield  {author} {\bibinfo {author} {\bibfnamefont {K.-M.~C.}\ \bibnamefont {Fu}}, \bibinfo {author} {\bibfnamefont {G.~Z.}\ \bibnamefont {Iwata}}, \bibinfo {author} {\bibfnamefont {A.}~\bibnamefont {Wickenbrock}},\ and\ \bibinfo {author} {\bibfnamefont {D.}~\bibnamefont {Budker}},\ }\bibfield  {title} {\bibinfo {title} {Sensitive magnetometry in challenging environments},\ }\href@noop {} {\bibfield  {journal} {\bibinfo  {journal} {AVS Quantum Science}\ }\textbf {\bibinfo {volume} {2}} (\bibinfo {year} {2020})}\BibitemShut {NoStop}%
\bibitem [{\citenamefont {Wang}\ \emph {et~al.}(2015)\citenamefont {Wang}, \citenamefont {Feng}, \citenamefont {Zhang}, \citenamefont {Chen}, \citenamefont {Zheng}, \citenamefont {Guo}, \citenamefont {Zhang}, \citenamefont {Song}, \citenamefont {Guo}, \citenamefont {Fan} \emph {et~al.}}]{wang2015high}%
  \BibitemOpen
  \bibfield  {author} {\bibinfo {author} {\bibfnamefont {J.}~\bibnamefont {Wang}}, \bibinfo {author} {\bibfnamefont {F.}~\bibnamefont {Feng}}, \bibinfo {author} {\bibfnamefont {J.}~\bibnamefont {Zhang}}, \bibinfo {author} {\bibfnamefont {J.}~\bibnamefont {Chen}}, \bibinfo {author} {\bibfnamefont {Z.}~\bibnamefont {Zheng}}, \bibinfo {author} {\bibfnamefont {L.}~\bibnamefont {Guo}}, \bibinfo {author} {\bibfnamefont {W.}~\bibnamefont {Zhang}}, \bibinfo {author} {\bibfnamefont {X.}~\bibnamefont {Song}}, \bibinfo {author} {\bibfnamefont {G.}~\bibnamefont {Guo}}, \bibinfo {author} {\bibfnamefont {L.}~\bibnamefont {Fan}}, \emph {et~al.},\ }\bibfield  {title} {\bibinfo {title} {High-sensitivity temperature sensing using an implanted single nitrogen-vacancy center array in diamond},\ }\href@noop {} {\bibfield  {journal} {\bibinfo  {journal} {Physical Review B}\ }\textbf {\bibinfo {volume} {91}},\ \bibinfo {pages} {155404} (\bibinfo {year} {2015})}\BibitemShut {NoStop}%
\bibitem [{\citenamefont {Ho}\ \emph {et~al.}(2021)\citenamefont {Ho}, \citenamefont {Wong}, \citenamefont {Leung}, \citenamefont {Pang}, \citenamefont {Leung}, \citenamefont {Yip}, \citenamefont {Zhang}, \citenamefont {Xie}, \citenamefont {Goh},\ and\ \citenamefont {Yang}}]{ho2021recent}%
  \BibitemOpen
  \bibfield  {author} {\bibinfo {author} {\bibfnamefont {K.~O.}\ \bibnamefont {Ho}}, \bibinfo {author} {\bibfnamefont {K.~C.}\ \bibnamefont {Wong}}, \bibinfo {author} {\bibfnamefont {M.~Y.}\ \bibnamefont {Leung}}, \bibinfo {author} {\bibfnamefont {Y.~Y.}\ \bibnamefont {Pang}}, \bibinfo {author} {\bibfnamefont {W.~K.}\ \bibnamefont {Leung}}, \bibinfo {author} {\bibfnamefont {K.~Y.}\ \bibnamefont {Yip}}, \bibinfo {author} {\bibfnamefont {W.}~\bibnamefont {Zhang}}, \bibinfo {author} {\bibfnamefont {J.}~\bibnamefont {Xie}}, \bibinfo {author} {\bibfnamefont {S.~K.}\ \bibnamefont {Goh}},\ and\ \bibinfo {author} {\bibfnamefont {S.}~\bibnamefont {Yang}},\ }\bibfield  {title} {\bibinfo {title} {Recent developments of quantum sensing under pressurized environment using the nitrogen vacancy (nv) center in diamond},\ }\href@noop {} {\bibfield  {journal} {\bibinfo  {journal} {Journal of Applied Physics}\ }\textbf {\bibinfo {volume} {129}} (\bibinfo {year} {2021})}\BibitemShut {NoStop}%
\bibitem [{\citenamefont {Kuwahata}\ \emph {et~al.}(2020)\citenamefont {Kuwahata}, \citenamefont {Kitaizumi}, \citenamefont {Saichi}, \citenamefont {Sato}, \citenamefont {Igarashi}, \citenamefont {Ohshima}, \citenamefont {Masuyama}, \citenamefont {Iwasaki}, \citenamefont {Hatano}, \citenamefont {Jelezko} \emph {et~al.}}]{kuwahata2020magnetometer}%
  \BibitemOpen
  \bibfield  {author} {\bibinfo {author} {\bibfnamefont {A.}~\bibnamefont {Kuwahata}}, \bibinfo {author} {\bibfnamefont {T.}~\bibnamefont {Kitaizumi}}, \bibinfo {author} {\bibfnamefont {K.}~\bibnamefont {Saichi}}, \bibinfo {author} {\bibfnamefont {T.}~\bibnamefont {Sato}}, \bibinfo {author} {\bibfnamefont {R.}~\bibnamefont {Igarashi}}, \bibinfo {author} {\bibfnamefont {T.}~\bibnamefont {Ohshima}}, \bibinfo {author} {\bibfnamefont {Y.}~\bibnamefont {Masuyama}}, \bibinfo {author} {\bibfnamefont {T.}~\bibnamefont {Iwasaki}}, \bibinfo {author} {\bibfnamefont {M.}~\bibnamefont {Hatano}}, \bibinfo {author} {\bibfnamefont {F.}~\bibnamefont {Jelezko}}, \emph {et~al.},\ }\bibfield  {title} {\bibinfo {title} {Magnetometer with nitrogen-vacancy center in a bulk diamond for detecting magnetic nanoparticles in biomedical applications},\ }\href@noop {} {\bibfield  {journal} {\bibinfo  {journal} {Scientific reports}\ }\textbf {\bibinfo {volume} {10}},\ \bibinfo {pages} {2483} (\bibinfo {year} {2020})}\BibitemShut {NoStop}%
\bibitem [{\citenamefont {Dolde}\ \emph {et~al.}(2011)\citenamefont {Dolde}, \citenamefont {Fedder}, \citenamefont {Doherty}, \citenamefont {N{\"o}bauer}, \citenamefont {Rempp}, \citenamefont {Balasubramanian}, \citenamefont {Wolf}, \citenamefont {Reinhard}, \citenamefont {Hollenberg}, \citenamefont {Jelezko} \emph {et~al.}}]{dolde2011electric}%
  \BibitemOpen
  \bibfield  {author} {\bibinfo {author} {\bibfnamefont {F.}~\bibnamefont {Dolde}}, \bibinfo {author} {\bibfnamefont {H.}~\bibnamefont {Fedder}}, \bibinfo {author} {\bibfnamefont {M.~W.}\ \bibnamefont {Doherty}}, \bibinfo {author} {\bibfnamefont {T.}~\bibnamefont {N{\"o}bauer}}, \bibinfo {author} {\bibfnamefont {F.}~\bibnamefont {Rempp}}, \bibinfo {author} {\bibfnamefont {G.}~\bibnamefont {Balasubramanian}}, \bibinfo {author} {\bibfnamefont {T.}~\bibnamefont {Wolf}}, \bibinfo {author} {\bibfnamefont {F.}~\bibnamefont {Reinhard}}, \bibinfo {author} {\bibfnamefont {L.~C.}\ \bibnamefont {Hollenberg}}, \bibinfo {author} {\bibfnamefont {F.}~\bibnamefont {Jelezko}}, \emph {et~al.},\ }\bibfield  {title} {\bibinfo {title} {Electric-field sensing using single diamond spins},\ }\href {https://rdcu.be/c3wnW} {\bibfield  {journal} {\bibinfo  {journal} {Nature Physics}\ }\textbf {\bibinfo {volume} {7}},\ \bibinfo {pages} {459} (\bibinfo {year} {2011})}\BibitemShut {NoStop}%
\bibitem [{\citenamefont {Barry}\ \emph {et~al.}(2023)\citenamefont {Barry}, \citenamefont {Steinecker}, \citenamefont {Alsid}, \citenamefont {Majumder}, \citenamefont {Pham}, \citenamefont {O'Keefe},\ and\ \citenamefont {Braje}}]{barry2023sensitive}%
  \BibitemOpen
  \bibfield  {author} {\bibinfo {author} {\bibfnamefont {J.~F.}\ \bibnamefont {Barry}}, \bibinfo {author} {\bibfnamefont {M.~H.}\ \bibnamefont {Steinecker}}, \bibinfo {author} {\bibfnamefont {S.~T.}\ \bibnamefont {Alsid}}, \bibinfo {author} {\bibfnamefont {J.}~\bibnamefont {Majumder}}, \bibinfo {author} {\bibfnamefont {L.~M.}\ \bibnamefont {Pham}}, \bibinfo {author} {\bibfnamefont {M.~F.}\ \bibnamefont {O'Keefe}},\ and\ \bibinfo {author} {\bibfnamefont {D.~A.}\ \bibnamefont {Braje}},\ }\bibfield  {title} {\bibinfo {title} {Sensitive ac and dc magnetometry with nitrogen-vacancy center ensembles in diamond},\ }\href@noop {} {\bibfield  {journal} {\bibinfo  {journal} {arXiv preprint arXiv:2305.06269}\ } (\bibinfo {year} {2023})}\BibitemShut {NoStop}%
\bibitem [{\citenamefont {Barry}\ \emph {et~al.}(2016)\citenamefont {Barry}, \citenamefont {Turner}, \citenamefont {Schloss}, \citenamefont {Glenn}, \citenamefont {Song}, \citenamefont {Lukin}, \citenamefont {Park},\ and\ \citenamefont {Walsworth}}]{barry2016optical}%
  \BibitemOpen
  \bibfield  {author} {\bibinfo {author} {\bibfnamefont {J.~F.}\ \bibnamefont {Barry}}, \bibinfo {author} {\bibfnamefont {M.~J.}\ \bibnamefont {Turner}}, \bibinfo {author} {\bibfnamefont {J.~M.}\ \bibnamefont {Schloss}}, \bibinfo {author} {\bibfnamefont {D.~R.}\ \bibnamefont {Glenn}}, \bibinfo {author} {\bibfnamefont {Y.}~\bibnamefont {Song}}, \bibinfo {author} {\bibfnamefont {M.~D.}\ \bibnamefont {Lukin}}, \bibinfo {author} {\bibfnamefont {H.}~\bibnamefont {Park}},\ and\ \bibinfo {author} {\bibfnamefont {R.~L.}\ \bibnamefont {Walsworth}},\ }\bibfield  {title} {\bibinfo {title} {Optical magnetic detection of single-neuron action potentials using quantum defects in diamond},\ }\href@noop {} {\bibfield  {journal} {\bibinfo  {journal} {Proceedings of the National Academy of Sciences}\ }\textbf {\bibinfo {volume} {113}},\ \bibinfo {pages} {14133} (\bibinfo {year} {2016})}\BibitemShut {NoStop}%
\bibitem [{\citenamefont {Lenz}\ \emph {et~al.}(2021{\natexlab{a}})\citenamefont {Lenz}, \citenamefont {Chatzidrosos}, \citenamefont {Wang}, \citenamefont {Bougas}, \citenamefont {Dumeige}, \citenamefont {Wickenbrock}, \citenamefont {Kerber}, \citenamefont {Z\'azvorka}, \citenamefont {Kammerbauer}, \citenamefont {Kl\"aui}, \citenamefont {Kazi}, \citenamefont {Fu}, \citenamefont {Itoh}, \citenamefont {Watanabe},\ and\ \citenamefont {Budker}}]{LenzPRA2021}%
  \BibitemOpen
  \bibfield  {author} {\bibinfo {author} {\bibfnamefont {T.}~\bibnamefont {Lenz}}, \bibinfo {author} {\bibfnamefont {G.}~\bibnamefont {Chatzidrosos}}, \bibinfo {author} {\bibfnamefont {Z.}~\bibnamefont {Wang}}, \bibinfo {author} {\bibfnamefont {L.}~\bibnamefont {Bougas}}, \bibinfo {author} {\bibfnamefont {Y.}~\bibnamefont {Dumeige}}, \bibinfo {author} {\bibfnamefont {A.}~\bibnamefont {Wickenbrock}}, \bibinfo {author} {\bibfnamefont {N.}~\bibnamefont {Kerber}}, \bibinfo {author} {\bibfnamefont {J.}~\bibnamefont {Z\'azvorka}}, \bibinfo {author} {\bibfnamefont {F.}~\bibnamefont {Kammerbauer}}, \bibinfo {author} {\bibfnamefont {M.}~\bibnamefont {Kl\"aui}}, \bibinfo {author} {\bibfnamefont {Z.}~\bibnamefont {Kazi}}, \bibinfo {author} {\bibfnamefont {K.-M.~C.}\ \bibnamefont {Fu}}, \bibinfo {author} {\bibfnamefont {K.~M.}\ \bibnamefont {Itoh}}, \bibinfo {author} {\bibfnamefont {H.}~\bibnamefont {Watanabe}},\ and\ \bibinfo {author} {\bibfnamefont {D.}~\bibnamefont {Budker}},\ }\bibfield  {title} {\bibinfo {title}
  {Imaging topological spin structures using light-polarization and magnetic microscopy},\ }\href {https://doi.org/10.1103/PhysRevApplied.15.024040} {\bibfield  {journal} {\bibinfo  {journal} {Phys. Rev. Appl.}\ }\textbf {\bibinfo {volume} {15}},\ \bibinfo {pages} {024040} (\bibinfo {year} {2021}{\natexlab{a}})}\BibitemShut {NoStop}%
\bibitem [{\citenamefont {Nowodzinski}\ \emph {et~al.}(2015)\citenamefont {Nowodzinski}, \citenamefont {Chipaux}, \citenamefont {Toraille}, \citenamefont {Jacques}, \citenamefont {Roch},\ and\ \citenamefont {Debuisschert}}]{nowodzinski2015nitrogen}%
  \BibitemOpen
  \bibfield  {author} {\bibinfo {author} {\bibfnamefont {A.}~\bibnamefont {Nowodzinski}}, \bibinfo {author} {\bibfnamefont {M.}~\bibnamefont {Chipaux}}, \bibinfo {author} {\bibfnamefont {L.}~\bibnamefont {Toraille}}, \bibinfo {author} {\bibfnamefont {V.}~\bibnamefont {Jacques}}, \bibinfo {author} {\bibfnamefont {J.-F.}\ \bibnamefont {Roch}},\ and\ \bibinfo {author} {\bibfnamefont {T.}~\bibnamefont {Debuisschert}},\ }\bibfield  {title} {\bibinfo {title} {Nitrogen-vacancy centers in diamond for current imaging at the redistributive layer level of integrated circuits},\ }\href@noop {} {\bibfield  {journal} {\bibinfo  {journal} {Microelectronics Reliability}\ }\textbf {\bibinfo {volume} {55}},\ \bibinfo {pages} {1549} (\bibinfo {year} {2015})}\BibitemShut {NoStop}%
\bibitem [{\citenamefont {Levine}\ \emph {et~al.}(2019)\citenamefont {Levine}, \citenamefont {Turner}, \citenamefont {Kehayias}, \citenamefont {Hart}, \citenamefont {Langellier}, \citenamefont {Trubko}, \citenamefont {Glenn}, \citenamefont {Fu},\ and\ \citenamefont {Walsworth}}]{levine2019principles}%
  \BibitemOpen
  \bibfield  {author} {\bibinfo {author} {\bibfnamefont {E.~V.}\ \bibnamefont {Levine}}, \bibinfo {author} {\bibfnamefont {M.~J.}\ \bibnamefont {Turner}}, \bibinfo {author} {\bibfnamefont {P.}~\bibnamefont {Kehayias}}, \bibinfo {author} {\bibfnamefont {C.~A.}\ \bibnamefont {Hart}}, \bibinfo {author} {\bibfnamefont {N.}~\bibnamefont {Langellier}}, \bibinfo {author} {\bibfnamefont {R.}~\bibnamefont {Trubko}}, \bibinfo {author} {\bibfnamefont {D.~R.}\ \bibnamefont {Glenn}}, \bibinfo {author} {\bibfnamefont {R.~R.}\ \bibnamefont {Fu}},\ and\ \bibinfo {author} {\bibfnamefont {R.~L.}\ \bibnamefont {Walsworth}},\ }\bibfield  {title} {\bibinfo {title} {Principles and techniques of the quantum diamond microscope},\ }\href@noop {} {\bibfield  {journal} {\bibinfo  {journal} {Nanophotonics}\ }\textbf {\bibinfo {volume} {8}},\ \bibinfo {pages} {1945} (\bibinfo {year} {2019})}\BibitemShut {NoStop}%
\bibitem [{\citenamefont {Blanchard}\ \emph {et~al.}(2021)\citenamefont {Blanchard}, \citenamefont {Budker},\ and\ \citenamefont {Trabesinger}}]{BLANCHARD2021106886}%
  \BibitemOpen
  \bibfield  {author} {\bibinfo {author} {\bibfnamefont {J.~W.}\ \bibnamefont {Blanchard}}, \bibinfo {author} {\bibfnamefont {D.}~\bibnamefont {Budker}},\ and\ \bibinfo {author} {\bibfnamefont {A.}~\bibnamefont {Trabesinger}},\ }\bibfield  {title} {\bibinfo {title} {Lower than low: Perspectives on zero- to ultralow-field nuclear magnetic resonance},\ }\href {https://doi.org/https://doi.org/10.1016/j.jmr.2020.106886} {\bibfield  {journal} {\bibinfo  {journal} {Journal of Magnetic Resonance}\ }\textbf {\bibinfo {volume} {323}},\ \bibinfo {pages} {106886} (\bibinfo {year} {2021})}\BibitemShut {NoStop}%
\bibitem [{\citenamefont {Simpson}\ \emph {et~al.}(2016)\citenamefont {Simpson}, \citenamefont {Tetienne}, \citenamefont {McCoey}, \citenamefont {Ganesan}, \citenamefont {Hall}, \citenamefont {Petrou}, \citenamefont {Scholten},\ and\ \citenamefont {Hollenberg}}]{simpson2016magneto}%
  \BibitemOpen
  \bibfield  {author} {\bibinfo {author} {\bibfnamefont {D.~A.}\ \bibnamefont {Simpson}}, \bibinfo {author} {\bibfnamefont {J.-P.}\ \bibnamefont {Tetienne}}, \bibinfo {author} {\bibfnamefont {J.~M.}\ \bibnamefont {McCoey}}, \bibinfo {author} {\bibfnamefont {K.}~\bibnamefont {Ganesan}}, \bibinfo {author} {\bibfnamefont {L.~T.}\ \bibnamefont {Hall}}, \bibinfo {author} {\bibfnamefont {S.}~\bibnamefont {Petrou}}, \bibinfo {author} {\bibfnamefont {R.~E.}\ \bibnamefont {Scholten}},\ and\ \bibinfo {author} {\bibfnamefont {L.~C.}\ \bibnamefont {Hollenberg}},\ }\bibfield  {title} {\bibinfo {title} {Magneto-optical imaging of thin magnetic films using spins in diamond},\ }\href {https://doi.org/https://doi.org/10.1038/srep22797} {\bibfield  {journal} {\bibinfo  {journal} {Scientific reports}\ }\textbf {\bibinfo {volume} {6}},\ \bibinfo {pages} {1} (\bibinfo {year} {2016})}\BibitemShut {NoStop}%
\bibitem [{\citenamefont {Wickenbrock}\ \emph {et~al.}(2016)\citenamefont {Wickenbrock}, \citenamefont {Zheng}, \citenamefont {Bougas}, \citenamefont {Leefer}, \citenamefont {Afach}, \citenamefont {Jarmola}, \citenamefont {Acosta},\ and\ \citenamefont {Budker}}]{doi:10.1063/1.4960171}%
  \BibitemOpen
  \bibfield  {author} {\bibinfo {author} {\bibfnamefont {A.}~\bibnamefont {Wickenbrock}}, \bibinfo {author} {\bibfnamefont {H.}~\bibnamefont {Zheng}}, \bibinfo {author} {\bibfnamefont {L.}~\bibnamefont {Bougas}}, \bibinfo {author} {\bibfnamefont {N.}~\bibnamefont {Leefer}}, \bibinfo {author} {\bibfnamefont {S.}~\bibnamefont {Afach}}, \bibinfo {author} {\bibfnamefont {A.}~\bibnamefont {Jarmola}}, \bibinfo {author} {\bibfnamefont {V.~M.}\ \bibnamefont {Acosta}},\ and\ \bibinfo {author} {\bibfnamefont {D.}~\bibnamefont {Budker}},\ }\bibfield  {title} {\bibinfo {title} {Microwave-free magnetometry with nitrogen-vacancy centers in diamond},\ }\href {https://doi.org/10.1063/1.4960171} {\bibfield  {journal} {\bibinfo  {journal} {Applied Physics Letters}\ }\textbf {\bibinfo {volume} {109}},\ \bibinfo {pages} {053505} (\bibinfo {year} {2016})},\ \Eprint {https://arxiv.org/abs/https://doi.org/10.1063/1.4960171} {https://doi.org/10.1063/1.4960171} \BibitemShut {NoStop}%
\bibitem [{\citenamefont {Staacke}\ \emph {et~al.}(2020)\citenamefont {Staacke}, \citenamefont {John}, \citenamefont {Wunderlich}, \citenamefont {Horsthemke}, \citenamefont {Knolle}, \citenamefont {Laube}, \citenamefont {Glösekötter}, \citenamefont {Burchard}, \citenamefont {Abel},\ and\ \citenamefont {Meijer}}]{https://doi.org/10.1002/qute.202000037}%
  \BibitemOpen
  \bibfield  {author} {\bibinfo {author} {\bibfnamefont {R.}~\bibnamefont {Staacke}}, \bibinfo {author} {\bibfnamefont {R.}~\bibnamefont {John}}, \bibinfo {author} {\bibfnamefont {R.}~\bibnamefont {Wunderlich}}, \bibinfo {author} {\bibfnamefont {L.}~\bibnamefont {Horsthemke}}, \bibinfo {author} {\bibfnamefont {W.}~\bibnamefont {Knolle}}, \bibinfo {author} {\bibfnamefont {C.}~\bibnamefont {Laube}}, \bibinfo {author} {\bibfnamefont {P.}~\bibnamefont {Glösekötter}}, \bibinfo {author} {\bibfnamefont {B.}~\bibnamefont {Burchard}}, \bibinfo {author} {\bibfnamefont {B.}~\bibnamefont {Abel}},\ and\ \bibinfo {author} {\bibfnamefont {J.}~\bibnamefont {Meijer}},\ }\bibfield  {title} {\bibinfo {title} {Isotropic scalar quantum sensing of magnetic fields for industrial application},\ }\href {https://doi.org/https://doi.org/10.1002/qute.202000037} {\bibfield  {journal} {\bibinfo  {journal} {Advanced Quantum Technologies}\ }\textbf {\bibinfo {volume} {3}},\ \bibinfo {pages} {2000037} (\bibinfo {year} {2020})}\BibitemShut
  {NoStop}%
\bibitem [{\citenamefont {Wunderlich}\ \emph {et~al.}(2021)\citenamefont {Wunderlich}, \citenamefont {Staacke}, \citenamefont {Knolle}, \citenamefont {Abel},\ and\ \citenamefont {Meijer}}]{10.1063/5.0059330}%
  \BibitemOpen
  \bibfield  {author} {\bibinfo {author} {\bibfnamefont {R.}~\bibnamefont {Wunderlich}}, \bibinfo {author} {\bibfnamefont {R.}~\bibnamefont {Staacke}}, \bibinfo {author} {\bibfnamefont {W.}~\bibnamefont {Knolle}}, \bibinfo {author} {\bibfnamefont {B.}~\bibnamefont {Abel}},\ and\ \bibinfo {author} {\bibfnamefont {J.}~\bibnamefont {Meijer}},\ }\bibfield  {title} {\bibinfo {title} {{Magnetic field and angle-dependent photoluminescence of a fiber-coupled nitrogen vacancy rich diamond}},\ }\href {https://doi.org/10.1063/5.0059330} {\bibfield  {journal} {\bibinfo  {journal} {Journal of Applied Physics}\ }\textbf {\bibinfo {volume} {130}},\ \bibinfo {pages} {124901} (\bibinfo {year} {2021})},\ \Eprint {https://arxiv.org/abs/https://pubs.aip.org/aip/jap/article-pdf/doi/10.1063/5.0059330/13254249/124901\_1\_online.pdf} {https://pubs.aip.org/aip/jap/article-pdf/doi/10.1063/5.0059330/13254249/124901\_1\_online.pdf} \BibitemShut {NoStop}%
\bibitem [{\citenamefont {Zheng}\ \emph {et~al.}(2020)\citenamefont {Zheng}, \citenamefont {Sun}, \citenamefont {Chatzidrosos}, \citenamefont {Zhang}, \citenamefont {Nakamura}, \citenamefont {Sumiya}, \citenamefont {Ohshima}, \citenamefont {Isoya}, \citenamefont {Wrachtrup}, \citenamefont {Wickenbrock},\ and\ \citenamefont {Budker}}]{PhysRevApplied.13.044023}%
  \BibitemOpen
  \bibfield  {author} {\bibinfo {author} {\bibfnamefont {H.}~\bibnamefont {Zheng}}, \bibinfo {author} {\bibfnamefont {Z.}~\bibnamefont {Sun}}, \bibinfo {author} {\bibfnamefont {G.}~\bibnamefont {Chatzidrosos}}, \bibinfo {author} {\bibfnamefont {C.}~\bibnamefont {Zhang}}, \bibinfo {author} {\bibfnamefont {K.}~\bibnamefont {Nakamura}}, \bibinfo {author} {\bibfnamefont {H.}~\bibnamefont {Sumiya}}, \bibinfo {author} {\bibfnamefont {T.}~\bibnamefont {Ohshima}}, \bibinfo {author} {\bibfnamefont {J.}~\bibnamefont {Isoya}}, \bibinfo {author} {\bibfnamefont {J.}~\bibnamefont {Wrachtrup}}, \bibinfo {author} {\bibfnamefont {A.}~\bibnamefont {Wickenbrock}},\ and\ \bibinfo {author} {\bibfnamefont {D.}~\bibnamefont {Budker}},\ }\bibfield  {title} {\bibinfo {title} {Microwave-free vector magnetometry with nitrogen-vacancy centers along a single axis in diamond},\ }\href {https://doi.org/10.1103/PhysRevApplied.13.044023} {\bibfield  {journal} {\bibinfo  {journal} {Phys. Rev. Appl.}\ }\textbf {\bibinfo {volume} {13}},\
  \bibinfo {pages} {044023} (\bibinfo {year} {2020})}\BibitemShut {NoStop}%
\bibitem [{\citenamefont {Rebeirro}\ \emph {et~al.}(2024)\citenamefont {Rebeirro}, \citenamefont {Omar}, \citenamefont {Lenz}, \citenamefont {Dhungel}, \citenamefont {Bl{\"u}mler}, \citenamefont {Budker},\ and\ \citenamefont {Wickenbrock}}]{rebeirro2024microwave}%
  \BibitemOpen
  \bibfield  {author} {\bibinfo {author} {\bibfnamefont {J.~S.}\ \bibnamefont {Rebeirro}}, \bibinfo {author} {\bibfnamefont {M.}~\bibnamefont {Omar}}, \bibinfo {author} {\bibfnamefont {T.}~\bibnamefont {Lenz}}, \bibinfo {author} {\bibfnamefont {O.}~\bibnamefont {Dhungel}}, \bibinfo {author} {\bibfnamefont {P.}~\bibnamefont {Bl{\"u}mler}}, \bibinfo {author} {\bibfnamefont {D.}~\bibnamefont {Budker}},\ and\ \bibinfo {author} {\bibfnamefont {A.}~\bibnamefont {Wickenbrock}},\ }\bibfield  {title} {\bibinfo {title} {Microwave-free wide-field magnetometry using nitrogen-vacancy centers},\ }\href@noop {} {\bibfield  {journal} {\bibinfo  {journal} {Physical Review Applied}\ }\textbf {\bibinfo {volume} {21}},\ \bibinfo {pages} {044039} (\bibinfo {year} {2024})}\BibitemShut {NoStop}%
\bibitem [{\citenamefont {Rondin}\ \emph {et~al.}(2014)\citenamefont {Rondin}, \citenamefont {Tetienne}, \citenamefont {Hingant}, \citenamefont {Roch}, \citenamefont {Maletinsky},\ and\ \citenamefont {Jacques}}]{Rondin2014}%
  \BibitemOpen
  \bibfield  {author} {\bibinfo {author} {\bibfnamefont {L.}~\bibnamefont {Rondin}}, \bibinfo {author} {\bibfnamefont {J.-P.}\ \bibnamefont {Tetienne}}, \bibinfo {author} {\bibfnamefont {T.}~\bibnamefont {Hingant}}, \bibinfo {author} {\bibfnamefont {J.-F.}\ \bibnamefont {Roch}}, \bibinfo {author} {\bibfnamefont {P.}~\bibnamefont {Maletinsky}},\ and\ \bibinfo {author} {\bibfnamefont {V.}~\bibnamefont {Jacques}},\ }\bibfield  {title} {\bibinfo {title} {Magnetometry with nitrogen-vacancy defects in diamond},\ }\href {https://doi.org/10.1088/0034-4885/77/5/056503} {\bibfield  {journal} {\bibinfo  {journal} {Reports on Progress in Physics}\ }\textbf {\bibinfo {volume} {77}},\ \bibinfo {pages} {056503} (\bibinfo {year} {2014})}\BibitemShut {NoStop}%
\bibitem [{\citenamefont {Mr{\'o}zek}\ \emph {et~al.}(2015)\citenamefont {Mr{\'o}zek}, \citenamefont {Rudnicki}, \citenamefont {Kehayias}, \citenamefont {Jarmola}, \citenamefont {Budker},\ and\ \citenamefont {Gawlik}}]{mrozek2015EPJ}%
  \BibitemOpen
  \bibfield  {author} {\bibinfo {author} {\bibfnamefont {M.}~\bibnamefont {Mr{\'o}zek}}, \bibinfo {author} {\bibfnamefont {D.}~\bibnamefont {Rudnicki}}, \bibinfo {author} {\bibfnamefont {P.}~\bibnamefont {Kehayias}}, \bibinfo {author} {\bibfnamefont {A.}~\bibnamefont {Jarmola}}, \bibinfo {author} {\bibfnamefont {D.}~\bibnamefont {Budker}},\ and\ \bibinfo {author} {\bibfnamefont {W.}~\bibnamefont {Gawlik}},\ }\bibfield  {title} {\bibinfo {title} {Longitudinal spin relaxation in nitrogen-vacancy ensembles in diamond},\ }\href@noop {} {\bibfield  {journal} {\bibinfo  {journal} {EPJ Quantum Technology}\ }\textbf {\bibinfo {volume} {2}},\ \bibinfo {pages} {1} (\bibinfo {year} {2015})}\BibitemShut {NoStop}%
\bibitem [{\citenamefont {Blanchard}\ and\ \citenamefont {Budker}(2007)}]{blanchard2007zero}%
  \BibitemOpen
  \bibfield  {author} {\bibinfo {author} {\bibfnamefont {J.~W.}\ \bibnamefont {Blanchard}}\ and\ \bibinfo {author} {\bibfnamefont {D.}~\bibnamefont {Budker}},\ }\bibfield  {title} {\bibinfo {title} {Zero-to ultralow-field nmr},\ }\href@noop {} {\bibfield  {journal} {\bibinfo  {journal} {Emagres}\ ,\ \bibinfo {pages} {1395}} (\bibinfo {year} {2007})}\BibitemShut {NoStop}%
\bibitem [{\citenamefont {van Oort}\ and\ \citenamefont {Glasbeek}(1989)}]{Oort1989PRB}%
  \BibitemOpen
  \bibfield  {author} {\bibinfo {author} {\bibfnamefont {E.}~\bibnamefont {van Oort}}\ and\ \bibinfo {author} {\bibfnamefont {M.}~\bibnamefont {Glasbeek}},\ }\bibfield  {title} {\bibinfo {title} {Cross-relaxation dynamics of optically excited n-v centers in diamond},\ }\href {https://doi.org/10.1103/PhysRevB.40.6509} {\bibfield  {journal} {\bibinfo  {journal} {Phys. Rev. B}\ }\textbf {\bibinfo {volume} {40}},\ \bibinfo {pages} {6509} (\bibinfo {year} {1989})}\BibitemShut {NoStop}%
\bibitem [{\citenamefont {Kong}\ \emph {et~al.}(2018)\citenamefont {Kong}, \citenamefont {Zhao}, \citenamefont {Ye}, \citenamefont {Wang}, \citenamefont {Qin}, \citenamefont {Yu}, \citenamefont {Su}, \citenamefont {Shi},\ and\ \citenamefont {Du}}]{kong2018nanoscale}%
  \BibitemOpen
  \bibfield  {author} {\bibinfo {author} {\bibfnamefont {F.}~\bibnamefont {Kong}}, \bibinfo {author} {\bibfnamefont {P.}~\bibnamefont {Zhao}}, \bibinfo {author} {\bibfnamefont {X.}~\bibnamefont {Ye}}, \bibinfo {author} {\bibfnamefont {Z.}~\bibnamefont {Wang}}, \bibinfo {author} {\bibfnamefont {Z.}~\bibnamefont {Qin}}, \bibinfo {author} {\bibfnamefont {P.}~\bibnamefont {Yu}}, \bibinfo {author} {\bibfnamefont {J.}~\bibnamefont {Su}}, \bibinfo {author} {\bibfnamefont {F.}~\bibnamefont {Shi}},\ and\ \bibinfo {author} {\bibfnamefont {J.}~\bibnamefont {Du}},\ }\bibfield  {title} {\bibinfo {title} {Nanoscale zero-field electron spin resonance spectroscopy},\ }\href@noop {} {\bibfield  {journal} {\bibinfo  {journal} {Nature Communications}\ }\textbf {\bibinfo {volume} {9}},\ \bibinfo {pages} {1563} (\bibinfo {year} {2018})}\BibitemShut {NoStop}%
\bibitem [{\citenamefont {Wang}\ \emph {et~al.}(2022)\citenamefont {Wang}, \citenamefont {Liu}, \citenamefont {Fan}, \citenamefont {Feng}, \citenamefont {Leong}, \citenamefont {Finkler}, \citenamefont {Denisenko}, \citenamefont {Wrachtrup}, \citenamefont {Li},\ and\ \citenamefont {Liu}}]{wang2022zero}%
  \BibitemOpen
  \bibfield  {author} {\bibinfo {author} {\bibfnamefont {N.}~\bibnamefont {Wang}}, \bibinfo {author} {\bibfnamefont {C.-F.}\ \bibnamefont {Liu}}, \bibinfo {author} {\bibfnamefont {J.-W.}\ \bibnamefont {Fan}}, \bibinfo {author} {\bibfnamefont {X.}~\bibnamefont {Feng}}, \bibinfo {author} {\bibfnamefont {W.-H.}\ \bibnamefont {Leong}}, \bibinfo {author} {\bibfnamefont {A.}~\bibnamefont {Finkler}}, \bibinfo {author} {\bibfnamefont {A.}~\bibnamefont {Denisenko}}, \bibinfo {author} {\bibfnamefont {J.}~\bibnamefont {Wrachtrup}}, \bibinfo {author} {\bibfnamefont {Q.}~\bibnamefont {Li}},\ and\ \bibinfo {author} {\bibfnamefont {R.-B.}\ \bibnamefont {Liu}},\ }\bibfield  {title} {\bibinfo {title} {Zero-field magnetometry using hyperfine-biased nitrogen-vacancy centers near diamond surfaces},\ }\href@noop {} {\bibfield  {journal} {\bibinfo  {journal} {Physical Review Research}\ }\textbf {\bibinfo {volume} {4}},\ \bibinfo {pages} {013098} (\bibinfo {year} {2022})}\BibitemShut {NoStop}%
\bibitem [{\citenamefont {Zheng}\ \emph {et~al.}(2019)\citenamefont {Zheng}, \citenamefont {Xu}, \citenamefont {Iwata}, \citenamefont {Lenz}, \citenamefont {Michl}, \citenamefont {Yavkin}, \citenamefont {Nakamura}, \citenamefont {Sumiya}, \citenamefont {Ohshima}, \citenamefont {Isoya}, \citenamefont {Wrachtrup}, \citenamefont {Wickenbrock},\ and\ \citenamefont {Budker}}]{PhysRevApplied.11.064068}%
  \BibitemOpen
  \bibfield  {author} {\bibinfo {author} {\bibfnamefont {H.}~\bibnamefont {Zheng}}, \bibinfo {author} {\bibfnamefont {J.}~\bibnamefont {Xu}}, \bibinfo {author} {\bibfnamefont {G.~Z.}\ \bibnamefont {Iwata}}, \bibinfo {author} {\bibfnamefont {T.}~\bibnamefont {Lenz}}, \bibinfo {author} {\bibfnamefont {J.}~\bibnamefont {Michl}}, \bibinfo {author} {\bibfnamefont {B.}~\bibnamefont {Yavkin}}, \bibinfo {author} {\bibfnamefont {K.}~\bibnamefont {Nakamura}}, \bibinfo {author} {\bibfnamefont {H.}~\bibnamefont {Sumiya}}, \bibinfo {author} {\bibfnamefont {T.}~\bibnamefont {Ohshima}}, \bibinfo {author} {\bibfnamefont {J.}~\bibnamefont {Isoya}}, \bibinfo {author} {\bibfnamefont {J.}~\bibnamefont {Wrachtrup}}, \bibinfo {author} {\bibfnamefont {A.}~\bibnamefont {Wickenbrock}},\ and\ \bibinfo {author} {\bibfnamefont {D.}~\bibnamefont {Budker}},\ }\bibfield  {title} {\bibinfo {title} {Zero-field magnetometry based on nitrogen-vacancy ensembles in diamond},\ }\href {https://doi.org/10.1103/PhysRevApplied.11.064068} {\bibfield
  {journal} {\bibinfo  {journal} {Phys. Rev. Appl.}\ }\textbf {\bibinfo {volume} {11}},\ \bibinfo {pages} {064068} (\bibinfo {year} {2019})}\BibitemShut {NoStop}%
\bibitem [{\citenamefont {Lenz}\ \emph {et~al.}(2021{\natexlab{b}})\citenamefont {Lenz}, \citenamefont {Wickenbrock}, \citenamefont {Jelezko}, \citenamefont {Balasubramanian},\ and\ \citenamefont {Budker}}]{lenz2021}%
  \BibitemOpen
  \bibfield  {author} {\bibinfo {author} {\bibfnamefont {T.}~\bibnamefont {Lenz}}, \bibinfo {author} {\bibfnamefont {A.}~\bibnamefont {Wickenbrock}}, \bibinfo {author} {\bibfnamefont {F.}~\bibnamefont {Jelezko}}, \bibinfo {author} {\bibfnamefont {G.}~\bibnamefont {Balasubramanian}},\ and\ \bibinfo {author} {\bibfnamefont {D.}~\bibnamefont {Budker}},\ }\bibfield  {title} {\bibinfo {title} {Magnetic sensing at zero field with a single nitrogen-vacancy center},\ }\href {https://doi.org/10.1088/2058-9565/abffbd} {\bibfield  {journal} {\bibinfo  {journal} {Quantum Science and Technology}\ }\textbf {\bibinfo {volume} {6}},\ \bibinfo {pages} {034006} (\bibinfo {year} {2021}{\natexlab{b}})}\BibitemShut {NoStop}%
\bibitem [{\citenamefont {Clevenson}\ \emph {et~al.}(2016)\citenamefont {Clevenson}, \citenamefont {Chen}, \citenamefont {Dolde}, \citenamefont {Teale}, \citenamefont {Englund},\ and\ \citenamefont {Braje}}]{PhysRevA.94.021401}%
  \BibitemOpen
  \bibfield  {author} {\bibinfo {author} {\bibfnamefont {H.}~\bibnamefont {Clevenson}}, \bibinfo {author} {\bibfnamefont {E.~H.}\ \bibnamefont {Chen}}, \bibinfo {author} {\bibfnamefont {F.}~\bibnamefont {Dolde}}, \bibinfo {author} {\bibfnamefont {C.}~\bibnamefont {Teale}}, \bibinfo {author} {\bibfnamefont {D.}~\bibnamefont {Englund}},\ and\ \bibinfo {author} {\bibfnamefont {D.}~\bibnamefont {Braje}},\ }\bibfield  {title} {\bibinfo {title} {Diamond-nitrogen-vacancy electronic and nuclear spin-state anticrossings under weak transverse magnetic fields},\ }\href {https://doi.org/10.1103/PhysRevA.94.021401} {\bibfield  {journal} {\bibinfo  {journal} {Phys. Rev. A}\ }\textbf {\bibinfo {volume} {94}},\ \bibinfo {pages} {021401} (\bibinfo {year} {2016})}\BibitemShut {NoStop}%
\bibitem [{\citenamefont {Anishchik}\ \emph {et~al.}(2015)\citenamefont {Anishchik}, \citenamefont {Vins}, \citenamefont {Yelisseyev}, \citenamefont {Lukzen}, \citenamefont {Lavrik},\ and\ \citenamefont {Bagryansky}}]{Anishchik2015}%
  \BibitemOpen
  \bibfield  {author} {\bibinfo {author} {\bibfnamefont {S.~V.}\ \bibnamefont {Anishchik}}, \bibinfo {author} {\bibfnamefont {V.~G.}\ \bibnamefont {Vins}}, \bibinfo {author} {\bibfnamefont {A.~P.}\ \bibnamefont {Yelisseyev}}, \bibinfo {author} {\bibfnamefont {N.~N.}\ \bibnamefont {Lukzen}}, \bibinfo {author} {\bibfnamefont {N.~L.}\ \bibnamefont {Lavrik}},\ and\ \bibinfo {author} {\bibfnamefont {V.~A.}\ \bibnamefont {Bagryansky}},\ }\bibfield  {title} {\bibinfo {title} {Low-field feature in the magnetic spectra of nv-centers in diamond},\ }\href {https://doi.org/10.1088/1367-2630/17/2/023040} {\bibfield  {journal} {\bibinfo  {journal} {New Journal of Physics}\ }\textbf {\bibinfo {volume} {17}},\ \bibinfo {pages} {023040} (\bibinfo {year} {2015})}\BibitemShut {NoStop}%
\bibitem [{\citenamefont {Akhmedzhanov}\ \emph {et~al.}(2017)\citenamefont {Akhmedzhanov}, \citenamefont {Gushchin}, \citenamefont {Nizov}, \citenamefont {Nizov}, \citenamefont {Sobgayda}, \citenamefont {Zelensky},\ and\ \citenamefont {Hemmer}}]{AkhmedzhanovPRA2017}%
  \BibitemOpen
  \bibfield  {author} {\bibinfo {author} {\bibfnamefont {R.}~\bibnamefont {Akhmedzhanov}}, \bibinfo {author} {\bibfnamefont {L.}~\bibnamefont {Gushchin}}, \bibinfo {author} {\bibfnamefont {N.}~\bibnamefont {Nizov}}, \bibinfo {author} {\bibfnamefont {V.}~\bibnamefont {Nizov}}, \bibinfo {author} {\bibfnamefont {D.}~\bibnamefont {Sobgayda}}, \bibinfo {author} {\bibfnamefont {I.}~\bibnamefont {Zelensky}},\ and\ \bibinfo {author} {\bibfnamefont {P.}~\bibnamefont {Hemmer}},\ }\bibfield  {title} {\bibinfo {title} {Microwave-free magnetometry based on cross-relaxation resonances in diamond nitrogen-vacancy centers},\ }\href {https://doi.org/10.1103/PhysRevA.96.013806} {\bibfield  {journal} {\bibinfo  {journal} {Phys. Rev. A}\ }\textbf {\bibinfo {volume} {96}},\ \bibinfo {pages} {013806} (\bibinfo {year} {2017})}\BibitemShut {NoStop}%
\bibitem [{\citenamefont {Cambria}\ \emph {et~al.}(2023)\citenamefont {Cambria}, \citenamefont {Norambuena}, \citenamefont {Dinani}, \citenamefont {Thiering}, \citenamefont {Gardill}, \citenamefont {Kemeny}, \citenamefont {Li}, \citenamefont {Lordi}, \citenamefont {Gali}, \citenamefont {Maze} \emph {et~al.}}]{cambria2023temperature}%
  \BibitemOpen
  \bibfield  {author} {\bibinfo {author} {\bibfnamefont {M.}~\bibnamefont {Cambria}}, \bibinfo {author} {\bibfnamefont {A.}~\bibnamefont {Norambuena}}, \bibinfo {author} {\bibfnamefont {H.}~\bibnamefont {Dinani}}, \bibinfo {author} {\bibfnamefont {G.}~\bibnamefont {Thiering}}, \bibinfo {author} {\bibfnamefont {A.}~\bibnamefont {Gardill}}, \bibinfo {author} {\bibfnamefont {I.}~\bibnamefont {Kemeny}}, \bibinfo {author} {\bibfnamefont {Y.}~\bibnamefont {Li}}, \bibinfo {author} {\bibfnamefont {V.}~\bibnamefont {Lordi}}, \bibinfo {author} {\bibfnamefont {{\'A}.}~\bibnamefont {Gali}}, \bibinfo {author} {\bibfnamefont {J.}~\bibnamefont {Maze}}, \emph {et~al.},\ }\bibfield  {title} {\bibinfo {title} {Temperature-dependent spin-lattice relaxation of the nitrogen-vacancy spin triplet in diamond},\ }\href@noop {} {\bibfield  {journal} {\bibinfo  {journal} {Physical Review Letters}\ }\textbf {\bibinfo {volume} {130}},\ \bibinfo {pages} {256903} (\bibinfo {year} {2023})}\BibitemShut {NoStop}%
\bibitem [{\citenamefont {Choi}\ \emph {et~al.}(2017)\citenamefont {Choi}, \citenamefont {Choi}, \citenamefont {Kucsko}, \citenamefont {Maurer}, \citenamefont {Shields}, \citenamefont {Sumiya}, \citenamefont {Onoda}, \citenamefont {Isoya}, \citenamefont {Demler}, \citenamefont {Jelezko}, \citenamefont {Yao},\ and\ \citenamefont {Lukin}}]{PhysRevLett.118.093601}%
  \BibitemOpen
  \bibfield  {author} {\bibinfo {author} {\bibfnamefont {J.}~\bibnamefont {Choi}}, \bibinfo {author} {\bibfnamefont {S.}~\bibnamefont {Choi}}, \bibinfo {author} {\bibfnamefont {G.}~\bibnamefont {Kucsko}}, \bibinfo {author} {\bibfnamefont {P.~C.}\ \bibnamefont {Maurer}}, \bibinfo {author} {\bibfnamefont {B.~J.}\ \bibnamefont {Shields}}, \bibinfo {author} {\bibfnamefont {H.}~\bibnamefont {Sumiya}}, \bibinfo {author} {\bibfnamefont {S.}~\bibnamefont {Onoda}}, \bibinfo {author} {\bibfnamefont {J.}~\bibnamefont {Isoya}}, \bibinfo {author} {\bibfnamefont {E.}~\bibnamefont {Demler}}, \bibinfo {author} {\bibfnamefont {F.}~\bibnamefont {Jelezko}}, \bibinfo {author} {\bibfnamefont {N.~Y.}\ \bibnamefont {Yao}},\ and\ \bibinfo {author} {\bibfnamefont {M.~D.}\ \bibnamefont {Lukin}},\ }\bibfield  {title} {\bibinfo {title} {Depolarization dynamics in a strongly interacting solid-state spin ensemble},\ }\href {https://doi.org/10.1103/PhysRevLett.118.093601} {\bibfield  {journal} {\bibinfo  {journal} {Phys. Rev. Lett.}\
  }\textbf {\bibinfo {volume} {118}},\ \bibinfo {pages} {093601} (\bibinfo {year} {2017})}\BibitemShut {NoStop}%
\bibitem [{\citenamefont {Perdriat}\ \emph {et~al.}(2023)\citenamefont {Perdriat}, \citenamefont {Huillery},\ and\ \citenamefont {H{\'e}tet}}]{pellet2023relaxation}%
  \BibitemOpen
  \bibfield  {author} {\bibinfo {author} {\bibfnamefont {M.}~\bibnamefont {Perdriat}}, \bibinfo {author} {\bibfnamefont {P.}~\bibnamefont {Huillery}},\ and\ \bibinfo {author} {\bibfnamefont {G.}~\bibnamefont {H{\'e}tet}},\ }\bibfield  {title} {\bibinfo {title} {Relaxation processes in dipole-coupled nitrogen-vacancy centers in zero field: Application in magnetometry},\ }\href@noop {} {\bibfield  {journal} {\bibinfo  {journal} {Physical Review Applied}\ }\textbf {\bibinfo {volume} {20}},\ \bibinfo {pages} {034050} (\bibinfo {year} {2023})}\BibitemShut {NoStop}%
\bibitem [{\citenamefont {Dhungel}\ \emph {et~al.}(2024{\natexlab{a}})\citenamefont {Dhungel}, \citenamefont {Lenz}, \citenamefont {Omar}, \citenamefont {Rebeirro}, \citenamefont {Luu}, \citenamefont {Younesi}, \citenamefont {Ulbricht}, \citenamefont {Iv\'ady}, \citenamefont {Gali}, \citenamefont {Wickenbrock},\ and\ \citenamefont {Budker}}]{dhungel2023zero}%
  \BibitemOpen
  \bibfield  {author} {\bibinfo {author} {\bibfnamefont {O.}~\bibnamefont {Dhungel}}, \bibinfo {author} {\bibfnamefont {T.}~\bibnamefont {Lenz}}, \bibinfo {author} {\bibfnamefont {M.}~\bibnamefont {Omar}}, \bibinfo {author} {\bibfnamefont {J.~S.}\ \bibnamefont {Rebeirro}}, \bibinfo {author} {\bibfnamefont {M.-T.}\ \bibnamefont {Luu}}, \bibinfo {author} {\bibfnamefont {A.~T.}\ \bibnamefont {Younesi}}, \bibinfo {author} {\bibfnamefont {R.}~\bibnamefont {Ulbricht}}, \bibinfo {author} {\bibfnamefont {V.}~\bibnamefont {Iv\'ady}}, \bibinfo {author} {\bibfnamefont {A.}~\bibnamefont {Gali}}, \bibinfo {author} {\bibfnamefont {A.}~\bibnamefont {Wickenbrock}},\ and\ \bibinfo {author} {\bibfnamefont {D.}~\bibnamefont {Budker}},\ }\bibfield  {title} {\bibinfo {title} {Near zero-field microwave-free magnetometry with ensembles of nitrogen-vacancy centers in diamond},\ }\href {https://doi.org/10.1103/PhysRevB.109.224107} {\bibfield  {journal} {\bibinfo  {journal} {Phys. Rev. B}\ }\textbf {\bibinfo {volume} {109}},\ \bibinfo
  {pages} {224107} (\bibinfo {year} {2024}{\natexlab{a}})}\BibitemShut {NoStop}%
\bibitem [{\citenamefont {Dhungel}\ \emph {et~al.}(2024{\natexlab{b}})\citenamefont {Dhungel}, \citenamefont {Mr\'{o}zek}, \citenamefont {Lenz}, \citenamefont {Iv\'{a}dy}, \citenamefont {Gali}, \citenamefont {Wickenbrock}, \citenamefont {Budker}, \citenamefont {Gawlik},\ and\ \citenamefont {Wojciechowski}}]{Dhungel:24}%
  \BibitemOpen
  \bibfield  {author} {\bibinfo {author} {\bibfnamefont {O.}~\bibnamefont {Dhungel}}, \bibinfo {author} {\bibfnamefont {M.}~\bibnamefont {Mr\'{o}zek}}, \bibinfo {author} {\bibfnamefont {T.}~\bibnamefont {Lenz}}, \bibinfo {author} {\bibfnamefont {V.}~\bibnamefont {Iv\'{a}dy}}, \bibinfo {author} {\bibfnamefont {A.}~\bibnamefont {Gali}}, \bibinfo {author} {\bibfnamefont {A.}~\bibnamefont {Wickenbrock}}, \bibinfo {author} {\bibfnamefont {D.}~\bibnamefont {Budker}}, \bibinfo {author} {\bibfnamefont {W.}~\bibnamefont {Gawlik}},\ and\ \bibinfo {author} {\bibfnamefont {A.~M.}\ \bibnamefont {Wojciechowski}},\ }\bibfield  {title} {\bibinfo {title} {Near-zero-field microwave-free magnetometry with nitrogen-vacancy centers in nanodiamonds},\ }\href {https://doi.org/10.1364/OE.521124} {\bibfield  {journal} {\bibinfo  {journal} {Opt. Express}\ }\textbf {\bibinfo {volume} {32}},\ \bibinfo {pages} {21936} (\bibinfo {year} {2024}{\natexlab{b}})}\BibitemShut {NoStop}%
\bibitem [{\citenamefont {Huxter}\ \emph {et~al.}(2022)\citenamefont {Huxter}, \citenamefont {Palm}, \citenamefont {Davis}, \citenamefont {Welter}, \citenamefont {Lambert}, \citenamefont {Trassin},\ and\ \citenamefont {Degen}}]{huxter2022scanning}%
  \BibitemOpen
  \bibfield  {author} {\bibinfo {author} {\bibfnamefont {W.~S.}\ \bibnamefont {Huxter}}, \bibinfo {author} {\bibfnamefont {M.~L.}\ \bibnamefont {Palm}}, \bibinfo {author} {\bibfnamefont {M.~L.}\ \bibnamefont {Davis}}, \bibinfo {author} {\bibfnamefont {P.}~\bibnamefont {Welter}}, \bibinfo {author} {\bibfnamefont {C.-H.}\ \bibnamefont {Lambert}}, \bibinfo {author} {\bibfnamefont {M.}~\bibnamefont {Trassin}},\ and\ \bibinfo {author} {\bibfnamefont {C.~L.}\ \bibnamefont {Degen}},\ }\bibfield  {title} {\bibinfo {title} {Scanning gradiometry with a single spin quantum magnetometer},\ }\href@noop {} {\bibfield  {journal} {\bibinfo  {journal} {Nature Communications}\ }\textbf {\bibinfo {volume} {13}},\ \bibinfo {pages} {3761} (\bibinfo {year} {2022})}\BibitemShut {NoStop}%
\bibitem [{\citenamefont {Filipkowski}\ \emph {et~al.}(2022)\citenamefont {Filipkowski}, \citenamefont {Mr{\'o}zek}, \citenamefont {St{\k{e}}pniewski}, \citenamefont {Kierdaszuk}, \citenamefont {Drabi{\'n}ska}, \citenamefont {Karpate}, \citenamefont {G{\l}owacki}, \citenamefont {Ficek}, \citenamefont {Gawlik}, \citenamefont {Buczy{\'n}ski} \emph {et~al.}}]{filipkowski2022volumetric}%
  \BibitemOpen
  \bibfield  {author} {\bibinfo {author} {\bibfnamefont {A.}~\bibnamefont {Filipkowski}}, \bibinfo {author} {\bibfnamefont {M.}~\bibnamefont {Mr{\'o}zek}}, \bibinfo {author} {\bibfnamefont {G.}~\bibnamefont {St{\k{e}}pniewski}}, \bibinfo {author} {\bibfnamefont {J.}~\bibnamefont {Kierdaszuk}}, \bibinfo {author} {\bibfnamefont {A.}~\bibnamefont {Drabi{\'n}ska}}, \bibinfo {author} {\bibfnamefont {T.}~\bibnamefont {Karpate}}, \bibinfo {author} {\bibfnamefont {M.}~\bibnamefont {G{\l}owacki}}, \bibinfo {author} {\bibfnamefont {M.}~\bibnamefont {Ficek}}, \bibinfo {author} {\bibfnamefont {W.}~\bibnamefont {Gawlik}}, \bibinfo {author} {\bibfnamefont {R.}~\bibnamefont {Buczy{\'n}ski}}, \emph {et~al.},\ }\bibfield  {title} {\bibinfo {title} {Volumetric incorporation of nv diamond emitters in nanostructured f2 glass magneto-optical fiber probes},\ }\href@noop {} {\bibfield  {journal} {\bibinfo  {journal} {Carbon}\ }\textbf {\bibinfo {volume} {196}},\ \bibinfo {pages} {10} (\bibinfo {year} {2022})}\BibitemShut {NoStop}%
\bibitem [{\citenamefont {Mi}\ \emph {et~al.}(2021)\citenamefont {Mi}, \citenamefont {Chen}, \citenamefont {Tan}, \citenamefont {Dou}, \citenamefont {Yang}, \citenamefont {Turaga}, \citenamefont {Ren}, \citenamefont {Vajandar}, \citenamefont {Yuen}, \citenamefont {Osipowicz} \emph {et~al.}}]{mi2021quantifying}%
  \BibitemOpen
  \bibfield  {author} {\bibinfo {author} {\bibfnamefont {Z.}~\bibnamefont {Mi}}, \bibinfo {author} {\bibfnamefont {C.-B.}\ \bibnamefont {Chen}}, \bibinfo {author} {\bibfnamefont {H.~Q.}\ \bibnamefont {Tan}}, \bibinfo {author} {\bibfnamefont {Y.}~\bibnamefont {Dou}}, \bibinfo {author} {\bibfnamefont {C.}~\bibnamefont {Yang}}, \bibinfo {author} {\bibfnamefont {S.~P.}\ \bibnamefont {Turaga}}, \bibinfo {author} {\bibfnamefont {M.}~\bibnamefont {Ren}}, \bibinfo {author} {\bibfnamefont {S.~K.}\ \bibnamefont {Vajandar}}, \bibinfo {author} {\bibfnamefont {G.~H.}\ \bibnamefont {Yuen}}, \bibinfo {author} {\bibfnamefont {T.}~\bibnamefont {Osipowicz}}, \emph {et~al.},\ }\bibfield  {title} {\bibinfo {title} {Quantifying nanodiamonds biodistribution in whole cells with correlative iono-nanoscopy},\ }\href@noop {} {\bibfield  {journal} {\bibinfo  {journal} {Nature Communications}\ }\textbf {\bibinfo {volume} {12}},\ \bibinfo {pages} {4657} (\bibinfo {year} {2021})}\BibitemShut {NoStop}%
\bibitem [{\citenamefont {Chen}\ \emph {et~al.}(2022)\citenamefont {Chen}, \citenamefont {Lin}, \citenamefont {Cheng}, \citenamefont {Huang}, \citenamefont {Shao}, \citenamefont {Ye}, \citenamefont {Liu}, \citenamefont {Chen}, \citenamefont {Luo},\ and\ \citenamefont {Chen}}]{chen2022nanodiamond}%
  \BibitemOpen
  \bibfield  {author} {\bibinfo {author} {\bibfnamefont {Y.}~\bibnamefont {Chen}}, \bibinfo {author} {\bibfnamefont {Q.}~\bibnamefont {Lin}}, \bibinfo {author} {\bibfnamefont {H.}~\bibnamefont {Cheng}}, \bibinfo {author} {\bibfnamefont {H.}~\bibnamefont {Huang}}, \bibinfo {author} {\bibfnamefont {J.}~\bibnamefont {Shao}}, \bibinfo {author} {\bibfnamefont {Y.}~\bibnamefont {Ye}}, \bibinfo {author} {\bibfnamefont {G.-S.}\ \bibnamefont {Liu}}, \bibinfo {author} {\bibfnamefont {L.}~\bibnamefont {Chen}}, \bibinfo {author} {\bibfnamefont {Y.}~\bibnamefont {Luo}},\ and\ \bibinfo {author} {\bibfnamefont {Z.}~\bibnamefont {Chen}},\ }\bibfield  {title} {\bibinfo {title} {Nanodiamond-based optical-fiber quantum probe for magnetic field and biological sensing},\ }\href@noop {} {\bibfield  {journal} {\bibinfo  {journal} {ACS sensors}\ }\textbf {\bibinfo {volume} {7}},\ \bibinfo {pages} {3660} (\bibinfo {year} {2022})}\BibitemShut {NoStop}%
\bibitem [{\citenamefont {Sengottuvel}\ \emph {et~al.}(2022)\citenamefont {Sengottuvel}, \citenamefont {Mr{\'o}zek}, \citenamefont {Sawczak}, \citenamefont {G{\l}owacki}, \citenamefont {Ficek}, \citenamefont {Gawlik},\ and\ \citenamefont {Wojciechowski}}]{sengottuvel2022wide}%
  \BibitemOpen
  \bibfield  {author} {\bibinfo {author} {\bibfnamefont {S.}~\bibnamefont {Sengottuvel}}, \bibinfo {author} {\bibfnamefont {M.}~\bibnamefont {Mr{\'o}zek}}, \bibinfo {author} {\bibfnamefont {M.}~\bibnamefont {Sawczak}}, \bibinfo {author} {\bibfnamefont {M.~J.}\ \bibnamefont {G{\l}owacki}}, \bibinfo {author} {\bibfnamefont {M.}~\bibnamefont {Ficek}}, \bibinfo {author} {\bibfnamefont {W.}~\bibnamefont {Gawlik}},\ and\ \bibinfo {author} {\bibfnamefont {A.~M.}\ \bibnamefont {Wojciechowski}},\ }\bibfield  {title} {\bibinfo {title} {Wide-field magnetometry using nitrogen-vacancy color centers with randomly oriented micro-diamonds},\ }\href@noop {} {\bibfield  {journal} {\bibinfo  {journal} {Scientific Reports}\ }\textbf {\bibinfo {volume} {12}},\ \bibinfo {pages} {17997} (\bibinfo {year} {2022})}\BibitemShut {NoStop}%
\bibitem [{\citenamefont {DeVience}\ \emph {et~al.}(2015)\citenamefont {DeVience}, \citenamefont {Pham}, \citenamefont {Lovchinsky}, \citenamefont {Sushkov}, \citenamefont {Bar-Gill}, \citenamefont {Belthangady}, \citenamefont {Casola}, \citenamefont {Corbett}, \citenamefont {Zhang}, \citenamefont {Lukin} \emph {et~al.}}]{devience2015nanoscale}%
  \BibitemOpen
  \bibfield  {author} {\bibinfo {author} {\bibfnamefont {S.~J.}\ \bibnamefont {DeVience}}, \bibinfo {author} {\bibfnamefont {L.~M.}\ \bibnamefont {Pham}}, \bibinfo {author} {\bibfnamefont {I.}~\bibnamefont {Lovchinsky}}, \bibinfo {author} {\bibfnamefont {A.~O.}\ \bibnamefont {Sushkov}}, \bibinfo {author} {\bibfnamefont {N.}~\bibnamefont {Bar-Gill}}, \bibinfo {author} {\bibfnamefont {C.}~\bibnamefont {Belthangady}}, \bibinfo {author} {\bibfnamefont {F.}~\bibnamefont {Casola}}, \bibinfo {author} {\bibfnamefont {M.}~\bibnamefont {Corbett}}, \bibinfo {author} {\bibfnamefont {H.}~\bibnamefont {Zhang}}, \bibinfo {author} {\bibfnamefont {M.}~\bibnamefont {Lukin}}, \emph {et~al.},\ }\bibfield  {title} {\bibinfo {title} {Nanoscale nmr spectroscopy and imaging of multiple nuclear species},\ }\href@noop {} {\bibfield  {journal} {\bibinfo  {journal} {Nature nanotechnology}\ }\textbf {\bibinfo {volume} {10}},\ \bibinfo {pages} {129} (\bibinfo {year} {2015})}\BibitemShut {NoStop}%
\bibitem [{\citenamefont {Barbiero}\ \emph {et~al.}(2020)\citenamefont {Barbiero}, \citenamefont {Castelletto}, \citenamefont {Zhang}, \citenamefont {Chen}, \citenamefont {Charnley}, \citenamefont {Russell},\ and\ \citenamefont {Gu}}]{Barbiero20}%
  \BibitemOpen
  \bibfield  {author} {\bibinfo {author} {\bibfnamefont {M.}~\bibnamefont {Barbiero}}, \bibinfo {author} {\bibfnamefont {S.}~\bibnamefont {Castelletto}}, \bibinfo {author} {\bibfnamefont {Q.}~\bibnamefont {Zhang}}, \bibinfo {author} {\bibfnamefont {Y.}~\bibnamefont {Chen}}, \bibinfo {author} {\bibfnamefont {M.}~\bibnamefont {Charnley}}, \bibinfo {author} {\bibfnamefont {S.}~\bibnamefont {Russell}},\ and\ \bibinfo {author} {\bibfnamefont {M.}~\bibnamefont {Gu}},\ }\bibfield  {title} {\bibinfo {title} {Nanoscale magnetic imaging enabled by nitrogen vacancy centres in nanodiamonds labelled by iron–oxide nanoparticles},\ }\href {https://doi.org/10.1039/C9NR10701K} {\bibfield  {journal} {\bibinfo  {journal} {Nanoscale}\ }\textbf {\bibinfo {volume} {12}},\ \bibinfo {pages} {8847} (\bibinfo {year} {2020})}\BibitemShut {NoStop}%
\bibitem [{\citenamefont {Ortner}\ and\ \citenamefont {Bandeira}(2020)}]{Ortner2020}%
  \BibitemOpen
  \bibfield  {author} {\bibinfo {author} {\bibfnamefont {M.}~\bibnamefont {Ortner}}\ and\ \bibinfo {author} {\bibfnamefont {L.~G.~C.}\ \bibnamefont {Bandeira}},\ }\bibfield  {title} {\bibinfo {title} {Magpylib: A free python package for magnetic field computation},\ }\href {https://doi.org/https://doi.org/10.1016/j.softx.2020.100466} {\bibfield  {journal} {\bibinfo  {journal} {SoftwareX}\ }\textbf {\bibinfo {volume} {11}},\ \bibinfo {pages} {100466} (\bibinfo {year} {2020})}\BibitemShut {NoStop}%
\bibitem [{\citenamefont {Barry}\ \emph {et~al.}(2020)\citenamefont {Barry}, \citenamefont {Schloss}, \citenamefont {Bauch}, \citenamefont {Turner}, \citenamefont {Hart}, \citenamefont {Pham},\ and\ \citenamefont {Walsworth}}]{Barry2020aps}%
  \BibitemOpen
  \bibfield  {author} {\bibinfo {author} {\bibfnamefont {J.~F.}\ \bibnamefont {Barry}}, \bibinfo {author} {\bibfnamefont {J.~M.}\ \bibnamefont {Schloss}}, \bibinfo {author} {\bibfnamefont {E.}~\bibnamefont {Bauch}}, \bibinfo {author} {\bibfnamefont {M.~J.}\ \bibnamefont {Turner}}, \bibinfo {author} {\bibfnamefont {C.~A.}\ \bibnamefont {Hart}}, \bibinfo {author} {\bibfnamefont {L.~M.}\ \bibnamefont {Pham}},\ and\ \bibinfo {author} {\bibfnamefont {R.~L.}\ \bibnamefont {Walsworth}},\ }\bibfield  {title} {\bibinfo {title} {Sensitivity optimization for nv-diamond magnetometry},\ }\href {https://doi.org/10.1103/RevModPhys.92.015004} {\bibfield  {journal} {\bibinfo  {journal} {Rev. Mod. Phys.}\ }\textbf {\bibinfo {volume} {92}},\ \bibinfo {pages} {015004} (\bibinfo {year} {2020})}\BibitemShut {NoStop}%
\bibitem [{\citenamefont {Dr\'eau}\ \emph {et~al.}(2011)\citenamefont {Dr\'eau}, \citenamefont {Lesik}, \citenamefont {Rondin}, \citenamefont {Spinicelli}, \citenamefont {Arcizet}, \citenamefont {Roch},\ and\ \citenamefont {Jacques}}]{Dreau2011PhysRev}%
  \BibitemOpen
  \bibfield  {author} {\bibinfo {author} {\bibfnamefont {A.}~\bibnamefont {Dr\'eau}}, \bibinfo {author} {\bibfnamefont {M.}~\bibnamefont {Lesik}}, \bibinfo {author} {\bibfnamefont {L.}~\bibnamefont {Rondin}}, \bibinfo {author} {\bibfnamefont {P.}~\bibnamefont {Spinicelli}}, \bibinfo {author} {\bibfnamefont {O.}~\bibnamefont {Arcizet}}, \bibinfo {author} {\bibfnamefont {J.-F.}\ \bibnamefont {Roch}},\ and\ \bibinfo {author} {\bibfnamefont {V.}~\bibnamefont {Jacques}},\ }\bibfield  {title} {\bibinfo {title} {Avoiding power broadening in optically detected magnetic resonance of single nv defects for enhanced dc magnetic field sensitivity},\ }\href {https://doi.org/10.1103/PhysRevB.84.195204} {\bibfield  {journal} {\bibinfo  {journal} {Phys. Rev. B}\ }\textbf {\bibinfo {volume} {84}},\ \bibinfo {pages} {195204} (\bibinfo {year} {2011})}\BibitemShut {NoStop}%
\end{thebibliography}%
\end{document}